\documentclass[useAMS,usenatbib]{mn2e}
\usepackage[dvips]{graphicx}
\usepackage{amsmath}
\usepackage{textcomp}

\title[SINFONI Spectroscopy of Massive YSOs in N113]{Integral Field Spectroscopy of Massive Young Stellar Objects in the N113 H\,{\sc ii} Region in the Large Magellanic Cloud}
 \author[J.L. Ward, J. M. Oliveira, J.Th. van Loon, M. Sewilo]{J.L. Ward$^{1}$\thanks{E-mail:
j.l.ward@keele.ac.uk (JLW); j.oliveira@keele.ac.uk (JMO)}, J.M. Oliveira$^{1}$\footnotemark[1], J.Th. van Loon$^{1}$ and M. Sewi\l{}o$^{2,3}$\\
$^{1}$School of Physical and Geographical Sciences, Lennard-Jones Laboratories, Keele University, Keele, ST5 5BG, UK \\
$^{2}$Space Science Institute, 4750 Walnut Street, Suite 205, Boulder, CO 80301, USA \\
$^{3}$The John Hopkins University, Department of Physics and Astronomy, 366 Bloomberg Center, 6400N. Charles Street, \\
Baltimore, MD 21218, USA}

\makeatother 

\begin{document}

\date{Accepted 2015 October 19.  Received 2015 September 11; in original form 2015 May 8}

\pagerange{\pageref{firstpage}--\pageref{lastpage}} \pubyear{2002}

\maketitle

\label{firstpage}

\begin{abstract}
The \textit{Spitzer} SAGE survey has allowed the identification and analysis of significant samples of Young Stellar Object (YSO) candidates in the Large Magellanic Cloud (LMC). 
However the angular resolution of \textit{Spitzer} is relatively poor meaning that at the distance of the LMC, it is likely that many of the \textit{Spitzer} 
YSO candidates in fact contain multiple components.
We present high resolution \textit{K}-band integral field spectroscopic observations of the three most prominent massive YSO candidates in the
N113 H\,{\sc ii} region using VLT/SINFONI. We have identified six \textit{K}-band continuum sources 
within the three \textit{Spitzer} sources and we have mapped the morphology and velocity fields of extended line emission around these sources.
Br$\gamma$, He\,{\sc i} and H$_2$ emission is
found at the position of all six \textit{K}-band sources; we discuss whether the emission is associated with the continuum sources or whether it is 
ambient emission.
H$_2$ emission appears to be mostly ambient emission and no evidence of CO emission arising in the
discs of YSOs has been found. We have mapped the centroid velocities of extended Br$\gamma$ emission and He {\sc i} emission and found evidence
of two expanding compact H\,{\sc ii} regions. One source shows compact and strong H$_2$ emission suggestive of a molecular outflow.
The diversity of spectroscopic properties observed is interpreted in the context of a range of evolutionary stages associated with massive star formation.

\end{abstract}


\section{Introduction}

\begin{figure*}
\begin{minipage}{175mm}
 \begin{center}
  \includegraphics[width=0.98\linewidth]{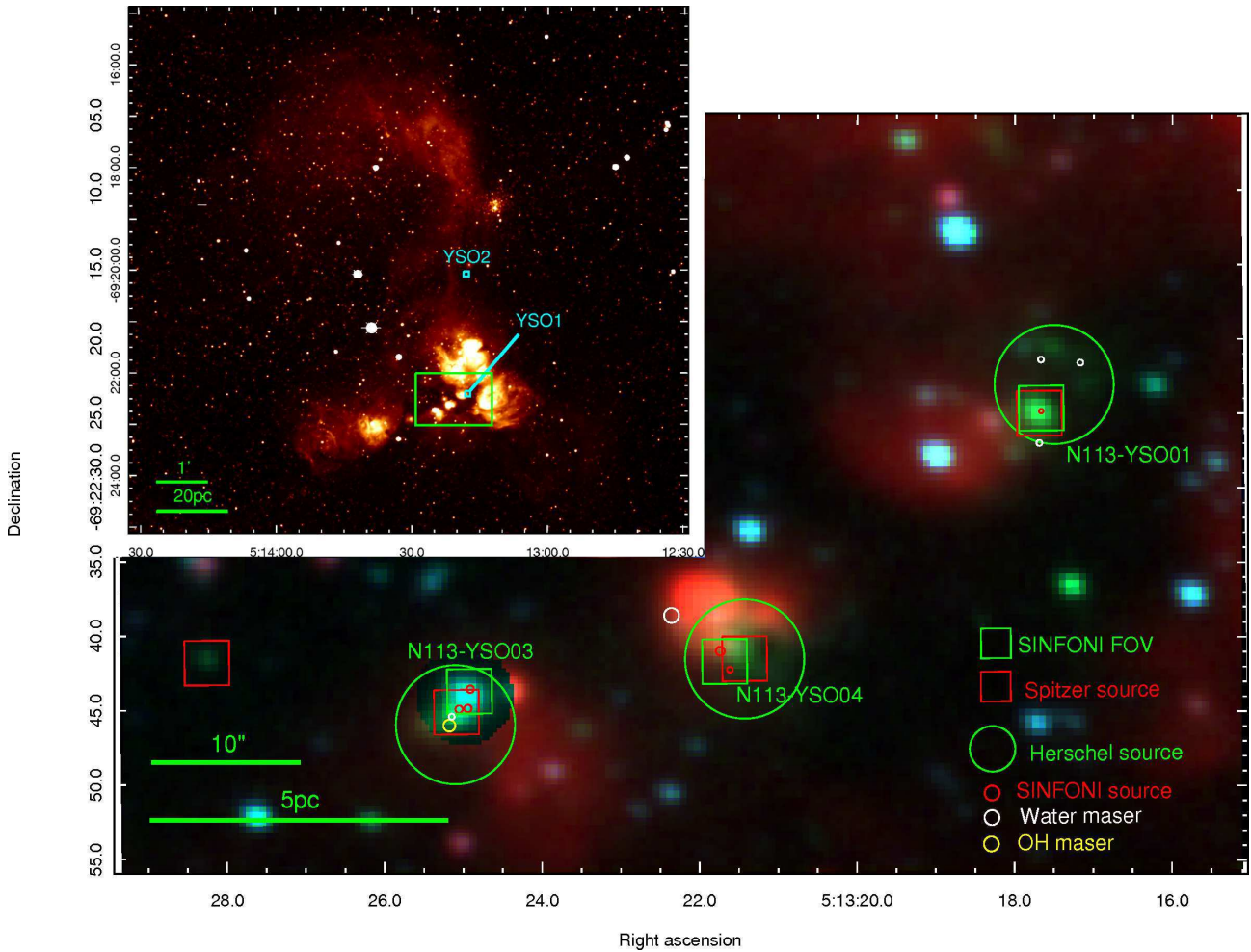}
\caption{Top left panel: wide field H$\alpha$ image$^1$ of N113 with positions of YSOs from \citet{Sewilo2010} (\textit{cyan squares}) and the region shown in the 
main image marked (\textit{green rectangle}).
Main image: three colour H$\alpha$ (red), IRSF \textit{Ks}-band (green) and IRSF \textit{H}-band (blue) \citep{Kato2007} composite showing positions of the observed N113 SINFONI FOVs (\textit{green squares}) and YSO candidates identified
in this work (\textit{red circles}).
Also included are the spectroscopically confirmed \textit{Spitzer} YSOs (\textit{red squares}), the \textit{Herschel} YSO candidates (\textit{green circles}; \citealt{Seale2014}), the water maser positions (\textit{white circles}) and the 
position of the OH maser (\textit{yellow circle}). The SINFONI \textit{red circles} represent the region from which a spectrum was extracted for each continuum source whilst 
all other symbol sizes are representative of the spatial accuracy of the data.}
 \end{center}
\end{minipage}
\end{figure*}

The Large Magellanic Cloud (LMC) presents a unique opportunity to study star formation. At a distance of $\sim$50 kpc \citep{Laney2012} and a favourable inclination, it allows
the simultaneous study of star formation on the scale of an entire galaxy and the scale of individual stars with little distance ambiguity. 
At the same time the LMC provides a \textquoteleft stepping stone' towards understanding star formation in lower metallicity 
environments with a metallicity of $Z_{\text{LMC}}\approx 0.4 \text{ Z}_{\odot}$ \citep{Dufour1982, Bernard2008}.

The \textit{Spitzer} Space Telescope (\textit{Spitzer}; \citealt{Werner2004}) and the \textit{Herschel} Space Observatory (\textit{Herschel}; \citealt{Pilbratt2010}) have allowed the identification
and characterisation of the stellar populations in the LMC through the  \textit{Spitzer} SAGE (\textquotedblleft Surveying the Agents of Galaxy Evolution"; \citealt{Meixner2006}) and 
the \textit{Herschel} HERITAGE (\textquotedblleft \textit{Herschel} Inventory of the Agents of Galaxy Evolution"; \citealt{Meixner2010, Meixner2013}) surveys.
These surveys covered the whole of the LMC at 3.6, 4.5, 5.8, 8.0, 24, 70, 160$\mu$m (SAGE) and at 100, 160, 250, 350, 500 $\mu$m (HERITAGE).
Over 2000 Young Stellar Object (YSO) candidates have been selected using \textit{Spitzer} photometry by \citet{Whitney2008}, \citet{Gruendl2009} and \citet{Carlson2012},
with spectroscopic analyses carried out in \citet{Shimonishi2008}, \citet{Seale2009}, \citet{Shimonishi2010}, \citet{Oliveira2009}, \citet{vanLoon2010LMC} and \citet{Woods2011}.
The spatial resolution of \textit{Spitzer} observations ranges from $1\rlap{.}^{\prime\prime}7$ at 3.6 $\mu$m to 40$^{\prime\prime}$ at 160 $\mu$m whilst 
those of \textit{Herschel} range from $8\rlap{.}^{\prime\prime}6$ (100 $\mu$m) to $40\rlap{.}^{\prime\prime}5$ (500 $\mu$m).
These resolutions are insufficient to distinguish between YSOs with separations less than $\sim$0.5 pc of
each other at the distance of the LMC.
The effect of this poor spatial resolution in \textit{Spitzer} imagery is that many \textit{Spitzer} sources classified as YSOs will in fact be multiple 
objects which may exhibit a wide range of characteristics.
 The physical properties of the sources based on the \textit{Spitzer} data may apply to groups of objects 
rather than individual sources.

LHA 120-N113 (hereafter N113, \citealt{Henize1956}) is an active star forming region within the LMC which contains a number of \textit{Spitzer} and \textit{Herschel} YSO candidate sources (see Fig. 1).
We present \textit{K}-band observations
of three of the brightest \textit{Spitzer} massive YSOs in N113
 obtained with SINFONI (Spectrograph for INtegral Field Observations in the Near Infrared; \citealt{Eisenhauer2003})  at the 
 European Southern Observatory (ESO) Very Large Telescope (VLT), with spectral and spatial resolutions of 4000 and 
 $0\rlap{.}^{\prime\prime}1$ respectively.
 \citet{Sewilo2010}
 discussed two \textit{Herschel} sources in N113 (YSO-1 and YSO-2, shown in Fig. 1) and two other YSOs (YSO-3 and YSO-4) in other regions of the LMC. 
To avoid confusion, we will retain the source numbers for YSO-1 and YSO-2 from \citet{Sewilo2010} but add the prefix N113- (N113-YSO01 and N113-YSO02) and 
add two additional YSOs in N113 (N113-YSO03 and N113-YSO04).

\footnotetext[1]{MOSAIC H$\alpha$ image of N113 from \textquotedblleft Magellanic Cloud Emission Line Survey 2" (PI: You-Hua Chu, http://adsabs.harvard.edu/abs/2011noao.prop..537C).}

In order to compare the massive YSO population of N113 with that of the Milky Way we require a suitable Galactic data set.
The Red MSX\footnote[2]{Midcourse Space Experiment \citep{Egan2003}.} Source Survey (RMS Survey; \citealt{Lumsden2013}) provides the most comprehensive catalogue of Galactic massive YSOs and 
Ultra-compact H\,{\sc ii} regions to date. \textit{H}- and \textit{K}-band infrared spectroscopy has been carried out for a large number of these objects 
by \citeauthor{Cooper2013} (2013, henceforth C13). This Galactic sample will be used as a comparison for our LMC YSO observations.

Whilst \textit{Spitzer} and \textit{Herschel} observations provide a valuable insight into star formation in the Magellanic Clouds, 
shorter wavelength studies in the near-infrared using large ground based
telescopes allow us to resolve individual hot cores and compact H\,{\sc ii} regions at the distance of the LMC.
These new observations provide the 
highest resolution imaging of these objects to date and the first \textit{K}-band spectroscopy of massive YSOs in this region. Parameters determined from 
previous observations are presented in Section 2.
Section 3 describes the observations and data reduction process, Section 4 presents the main results and finally, the significance and implications
of the results are discussed in Section 5, along with efforts to place the observed targets into an evolutionary context.

\section{Previous Observations}

\begin{table*}
 \begin{minipage}{175mm}
 \begin{center}
\caption{
Properties of \textit{Spitzer} YSOs analysed in this paper.
YSO02 has been included for completeness using the values from \citet{Sewilo2010}. The value for envelope mass is not included in \citet{Sewilo2010}.
The S09 group refers to the YSO classifications by Seale et al. (2009; S09) where P type sources show prominent PAH emission, PE sources show strong PAH and 
fine-structure emission. All three of the targets observed in this work have been classed as definite YSOs by \citet{Gruendl2009}
and by \citet{Carlson2012}.
The bolometric luminosities and masses for N113-YSO01, N113-YSO03 and N113-YSO04 are from new SED fits using existing photometry (see text for full details).
}
  \begin{tabular}{c c c c c c c}
\hline
&	&	&	&	&	&	Associated \\
   Target & \textit{Spitzer} Source & S09 group &  log(L$_{bol}$/L$_{\odot}$) & Central Mass (M$_{\odot}$) & Envelope Mass (M$_{\odot}$) &  Maser emission \\
   \hline
   N113-YSO02 &  \null &  \null & $4.51\substack{+0.29 \\ -0.30}$ & 13$\pm$2 & \null &  \null \\
  \hline
   N113-YSO01 & Y051317.69$-$692225.0 & PE & $5.18\substack{+0.18 \\ -0.08}$ & $32.3\substack{+0.3 \\ -0.7}$ & $1.8\substack{+6.3 \\ -0.2}$ $\times$10$^{2}$  & H$_2$O \\
   N113-YSO03 & Y051325.09$-$692245.1 & P & $5.27\substack{+0.28 \\ -0.27}$ & $35.3\substack{+12.8 \\ -0.9}$ & $1.6\substack{+2.3 \\ -1.1}$ $\times$10$^{2}$ & H$_2$O, OH \\
   N113-YSO04 & Y051321.43$-$692241.5 & PE & $5.24\substack{+0.31 \\ -0.51}$ & $34.3\substack{+13.9 \\ -15.8}$ & $1.2\substack{+33.3 \\ -0.6}$ $\times$10$^{2}$ & H$_2$O \\
\hline
\end{tabular}

 \end{center}

 \end{minipage}

\end{table*}

All three target regions (N113-YSO01, N113-YSO03 and N113-YSO04) are associated with bright knots in H$\alpha$ emission and have been previously studied using 
\textit{Spitzer} photometry \citep{Gruendl2009, Carlson2012}, \textit{Spitzer} IRS spectroscopy
\citep{Seale2009}, \textit{Herschel} photometry \citep{Sewilo2010} and \textit{Herschel} spectroscopy (Oliveira et al., in prep.), confirming the YSO classification.
Additionally, the region contains
a number of water masers including the most intense in the Magellanic Clouds \citep{Lazendic2002, Oliveira2006, Imai2013} and an OH maser \citep{Brooks1997}.
None of the targets in this paper appear in the YSO catalogue of \citet{Whitney2008} because they do not satisfy the strict point source criterion of the original SAGE point
source catalogue of \citet{Meixner2006}; since the sources are absent from the SAGE catalogue they are not part of the YSO selection.
This is most likely due to their slightly irregular morphologies and the issue is discussed in detail in \citet{Chen2009}.

The following spectroscopic properties are common to all three targets: H$_2$ emission, PAH emission and fine structure emission (from \textit{Spitzer}-IRS spectra, \citealt{Seale2009}), and [C\,{\sc ii}] emission, [O\,{\sc i}] 
emission, and CO emission (from \textit{Herschel} PACS and SPIRE spectroscopy, Oliveira et al. in prep.) whilst evidence of ices has not been observed in any of the targets. 
HCN and HCO$^{+}$ are also detected towards all three sources, indicating high densities (HCN and HCO$^{+}$) and photo-dissociation of molecular clouds (HCO$^{+}$),
both of which are associated with massive star formation \citep{Seale2012}. 

Additional properties,
including results of Spectral Energy Distribution (SED) fittting 
using the models of \citet{Robitaille2006}, are summarised in  Table 1. 
The SED fits for N113-YSO01, N113-YSO03 and N113-YSO04 are performed excluding the \textit{Herschel} photometry and setting the \textit{Spitzer} 70 $\mu$m data as upper limits 
because the spatial regions from which these data are extracted are too large to be directly useful when analysing individual YSOs. 
Physical parameters are estimated by averaging parameters of all models that fit a source's SED with normalised $\chi^2$ per data point $(\chi^2/pt)$ in a range 
between $(\chi^2/pt)_{\text{best}}$ for the best-fitting model and $(\chi^2/pt)_{\text{best}}+3$ (see e.g. \citealt{Sewilo2013} for further details). The luminosity and mass of N113-YSO01 are 
consistent within the quoted uncertainties with those presented in Sewilo et al. (2010), fitted using all available photometry for $\lambda < 50$ $\mu$m. Using the calibration 
from \citet{Martins2005} the estimated luminosities would correspond to O4--O9 stars on the main-sequence; however such estimates cannot be taken at face value 
since at least two sources are actually resolved into multiple components in the \textit{K}-band in this work. Whilst all three of the targets of this paper appear
similar based on \textit{Spitzer} and \textit{Herschel} data, the \textit{K}-band observations presented in this work reveal a wide range of sources with different
spectral properties.

\section{Observations and Data Reduction}

\textit{K}-band integral field spectroscopic observations were carried out for 3 targets in N113 using SINFONI
at the VLT. 
SINFONI is an AO-assisted integral field spectrograph allowing for observations in the \textit{K}-band with an angular 
resolution of $0\rlap{.}^{\prime\prime}1$, a field of view (FOV) of 2$\rlap{.}^{\prime\prime}$8
and a spectral resolving power, $R = \frac{\lambda}{\delta\lambda} = 4000$.
These observations took place in October 2013.
Each object was observed with four 300 second integrations along with sky offset position observations in 
an ABBA pattern with jittering. Telluric B-type standard stars were also observed at regular intervals throughout each night in order to provide standard star
spectra for telluric correction and flux calibration. Calibration frames were observed during the daytime and linearity lamp frames were obtained from the ESO archive.

The data were reduced using the standard SINFONI pipeline recipes with ESO's \textit{Gasgano} data file organiser. 
Telluric and flux calibration were performed simultaneously for each cube using an IDL script written specifically for this task. 
For each pixel in the data cube, the target spectrum is
 divided by the telluric standard star spectrum removing the telluric absorption features. 
The target spectrum was then multiplied by a blackbody with a temperature based on the spectral type of the standard star used. The blackbody spectra used in 
this calibration were generated using PyRAF\footnote[3]{PyRAF is a product of the Space Telescope Science Institute, which is operated by AURA for NASA.}. 
This process was looped to apply the same procedure to each spaxel in the cube.

SINFONI is a Cassegrain focus mounted instrument and as such it does suffer from a systematic
time-dependent wavelength shift during each night due to flexure. This is always small 
(less than 3 resolution elements)
so it only presents an issue when determinining accurate centroid velocity measurements. In order to 
account for this effect, a second wavelength calibration was performed on the final data cubes 
using the OH emission lines in the sky data cubes produced in the SINFONI data reduction pipeline.

Although sky line subtraction does form part of the standard SINFONI data reduction pipeline, 
due to the relatively long exposure times in this study (8 $\times$ 300 s exposures plus 
overheads), the 
variation in sky line intensities leads to sky line residuals remaining in the final data cubes.
The positions of these residuals are shown in Fig. A2. Whilst aesthetically displeasing, the impact of
these residuals on the spectral analysis is actually very small as none are coincident with any
emission lines of interest and the continuum measurements are calculated from models fitted to the 
continuum.

\section{Results}

\begin{figure*}
  \begin{minipage}{175mm}
 \begin{center}
\includegraphics[width=0.32\linewidth]{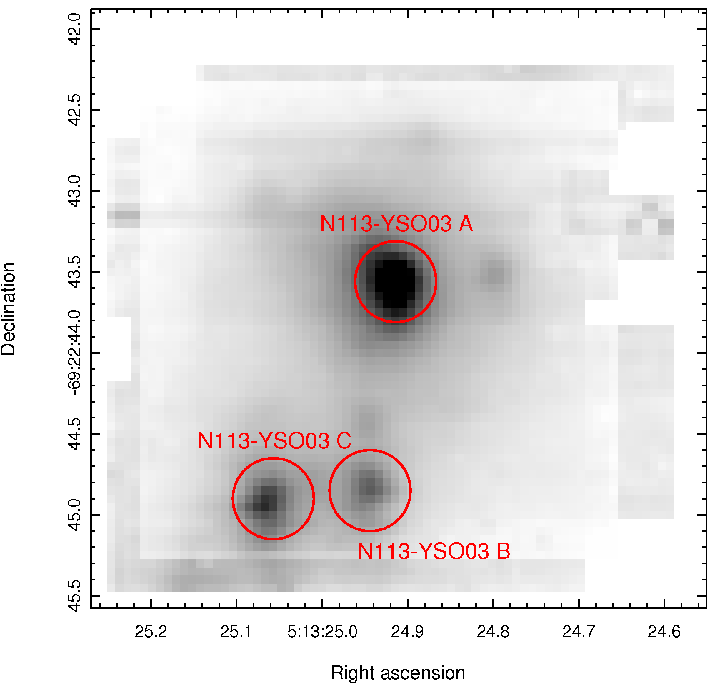}
\includegraphics[width=0.32\linewidth]{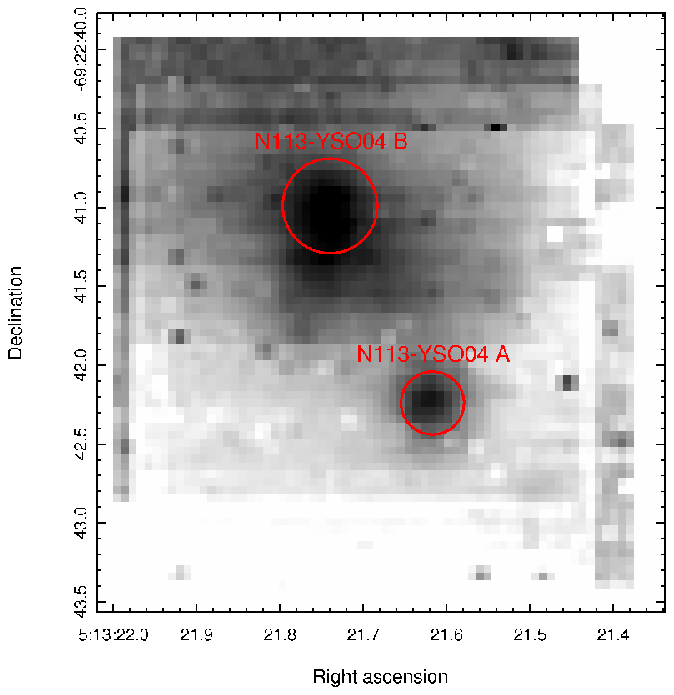}
  \includegraphics[width=0.32\linewidth]{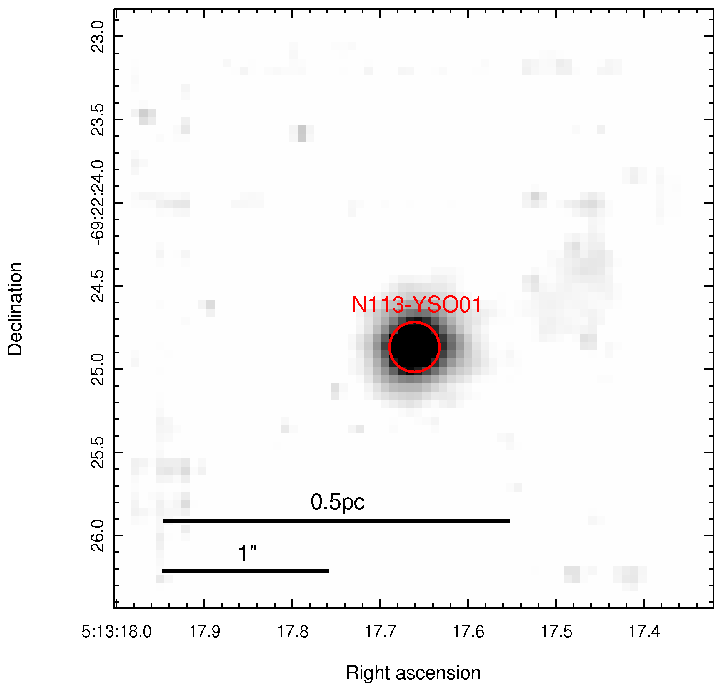}
\caption{SINFONI \textit{K}-band Continuum emission maps. Left to right: N113-YSO03, N113-YSO04, N113-YSO01. Marked regions show the identified continuum sources and the regions
from which spectra were extracted.}
 \end{center}
\end{minipage}
\end{figure*}

\subsection{Continuum emission and photometry}

For each spaxel in the final flux calibrated cubes, the continuum was fitted using a 3rd order polynomial and summed for the spectral range spanning 2.028--2.290 $\mu$m to produce
continuum flux maps without any contribution from line emission. The resulting images are shown in Fig. 2. Multiple continuum sources have been identified
using these images; N113-YSO01 contains a single continuum source while N113-YSO03 and N113-YSO4 are resolved into three and two continuum sources, respectively. 
The positions of each continuum source are given in Table 2 and marked in Fig. 2 and the red circles 
show the regions from which 1D spectra were extracted from the cubes using the \textit{sinfo\_utl\_cube2spectrum} recipe from the 
SINFONI data reduction pipeline (see Fig. A1 for extracted spectra).
The \textit{K}-band continuum magnitude 
for each object, integrated over the same wavelength interval, is given in Table 3.

\begin{table}
\caption{J2000 positions of each of the \textit{K}-band continuum sources resolved for the first time in this paper.}
\begin{center}
   \begin{tabular}{l l l}
\hline
    Object & RA & Dec \\
	   & (h:m:s) & (\textdegree:$^{\prime}$:$^{\prime\prime}$) \\
    \hline
    N113-YSO01  & 05:13:17.666 & $-$69:22:24.86 \\
    N113-YSO03 A & 05:13:24.915 & $-$69:22:43.55 \\
    N113-YSO03 B & 05:13:24.944 & $-$69:22:44.85 \\
    N113-YSO03 C & 05:13:25.057 & $-$69:22:44.90 \\
    N113-YSO04 A & 05:13:21.617 & $-$69:22:42.24 \\
    N113-YSO04 B & 05:13:21.740 & $-$69:22:40.99 \\
\hline
   \end{tabular}
\end{center}
\end{table}

\subsection{Extinction}

In order to impose constraints on the physical properties of the observed YSOs, we must apply extinction corrections to our measurements.
Additionally extinction towards an object can provide an assessment of how embedded the source of the emission is.
We employed two methods of estimating the extinction towards each source. The first using \textit{JHKs} photometry from the IRSF \citep{Kato2007} and 
the same technique employed in C13:
\begin{equation}
 A_{V} = \frac{m_{1}-m_{2}+x_{int}}{0.55^{1.75}(\lambda_{1}^{-1.75}-\lambda_{2}^{-1.75})}
\end{equation}
where $m_{1}$ and $m_{2}$ are the shorter wavelength and longer wavelength magnitudes respectively, and $x_{int}$ is the intrinsic colour,
assuming intrinsic colours of a B0 type star of 0.12 mag and 0.05 mag for \textit{J--H} and \textit{H--K}, respectively.

The second method utilises the H$_2$ line fluxes measured from the spectra themselves to estimate extinction. The 1-0(Q3) / 1-0(S1) flux ratio is used 
due to its insensitivity to temperature and relatively large wavelength baseline. Following \citet{Davis2011} A$_V$ is calculated as:
\begin{equation}
 A_{V} = -114\log(0.704[I_{S1}/I_{Q3}])
\end{equation}
All the resulting extinction estimates are shown in Table 3.
We find that the extinction estimates using the IRSF colours are inconsistent between \textit{J--H} and \textit{H--K} colours for the same source
 and for nearby objects in the same FOV.
Additionally the technique using IRSF photometry has yielded negative values, likely caused by source confusion and unreliable photometry in the relatively low 
resolution IRSF data or by the assumption of spectral type.
 Wherever extinction corrections are applied, we use the values calculated using the H$_2$ lines
as this technique makes no asssumptions of intrinsic spectral type and it is available for all six
sources. Using the Galactic mean $R_V$ dependent extinction law;
\begin{equation}
 [A(\lambda)/A_{V}] = a(x) + b(x)/R_{V}\text{,}
\end{equation}
where $a(x) = 0.574x^{1.61}$ and $b(x) = -0.527x^{1.61}$ for the \textit{K}-band \citep{Cardelli1989}, we calculated extinction corrections for all measured emission lines.
The $R_{V}$ value adopted is the same as that of the Milky Way extinction curve ($R_V = $ 3.1). Although 
the average value in the LMC has been found to be $R_V = $ 3.41$\pm$0.06 \citep{Gordon2003}, in the \textit{K}-band the effect of a small variation in
$R_V$ is negligible when compared to the effect of the line measurement uncertainties.

\begin{table*}

 \begin{minipage}{175mm}
\begin{center}
\caption{Measured \textit{K}-band magnitudes and extinction estimates (calculated using both methods discussed in Section 4.2) for all observed continuum sources. 
The average for each SINFONI FOV is given in the last column. For N113-YSO04 only one IRSF source is detected and it is unclear which of the sources in this work 
this corresponds to.}
\begin{tabular}{l l l l l l}
\hline
Target & \textit{K}-band mag & $A_V$ (\textit{J--H}) & $A_V$ (\textit{H--Ks}) & $A_V$ (H$_2$ 1-0 S(1)/Q(3)) & FOV average \\
\hline
N113-YSO01 & 16.27 $\pm$ 0.01 & \llap{1}0.7 $\pm$ 1.2 &  \llap{1}6.1 $\pm$ 1.1 & 24.1 $\pm$ 17.3 & 24.1 $\pm$ 17.3\\
N113-YSO03 A & 14.60 $\pm$ 0.03 & \llap{$-$}6.1 $\pm$ 2.4 &  \llap{$-$}1.6 $\pm$ 3.3 & 13.5 $\pm$ 8.5 & 15.0 $\pm$ 0.8 \\
N113-YSO03 B & 15.45 $\pm$ 0.03 & 1.8 $\pm$ 1.3 &  \llap{1}1.8 $\pm$ 1.0 & 15.7 $\pm$ 5.4 & 15.0 $\pm$ 0.8 \\
N113-YSO03 C & 15.20 $\pm$ 0.01 & 8.9 $\pm$ 1.4 & \null & 15.8 $\pm$ 3.9 & 15.0 $\pm$ 0.8 \\
N113-YSO04 A & 17.67 $\pm$ 0.03 & \null &  \llap{1}1.6 $\pm$ 1.6 & 37.8 $\pm$ 8.9 & 32.4 $\pm$ 5.5 \\
N113-YSO04 B & 16.53 $\pm$ 0.06 & \null &  \llap{1}1.6 $\pm$ 1.6 & 26.9 $\pm$ 6.5 & 32.4 $\pm$ 5.5 \\
\hline
\end{tabular}

\end{center}
\end{minipage}

\end{table*}

\subsection{Emission features}

Emission line mapping was achieved by fitting Gaussian profiles to the spectral axis in the final data cubes to
calculate a line flux for each spaxel using our IDL script written for this task. The resulting images are shown in Fig. 3.
Spectra were extracted from the regions shown in Fig. 2 of the flux calibrated data cubes using the \textit{sinfo\_utl\_cube2spectrum} recipe
from the SINFONI data reduction pipeline.
Emission lines in the extracted 1D spectra for each continuum source were measured by Gaussian fitting within the Starlink software package \textit{SPLAT}.
The measured emission line fluxes (without extinction correction) are given in Table B1.
The flux values obtained from the spectra were converted to line luminosities using a distance of 49.4$\pm$0.5 kpc \citep{Laney2012}.

\begin{figure*}
 \begin{minipage}{175mm}
 \begin{center}
\includegraphics[width=0.32\linewidth]{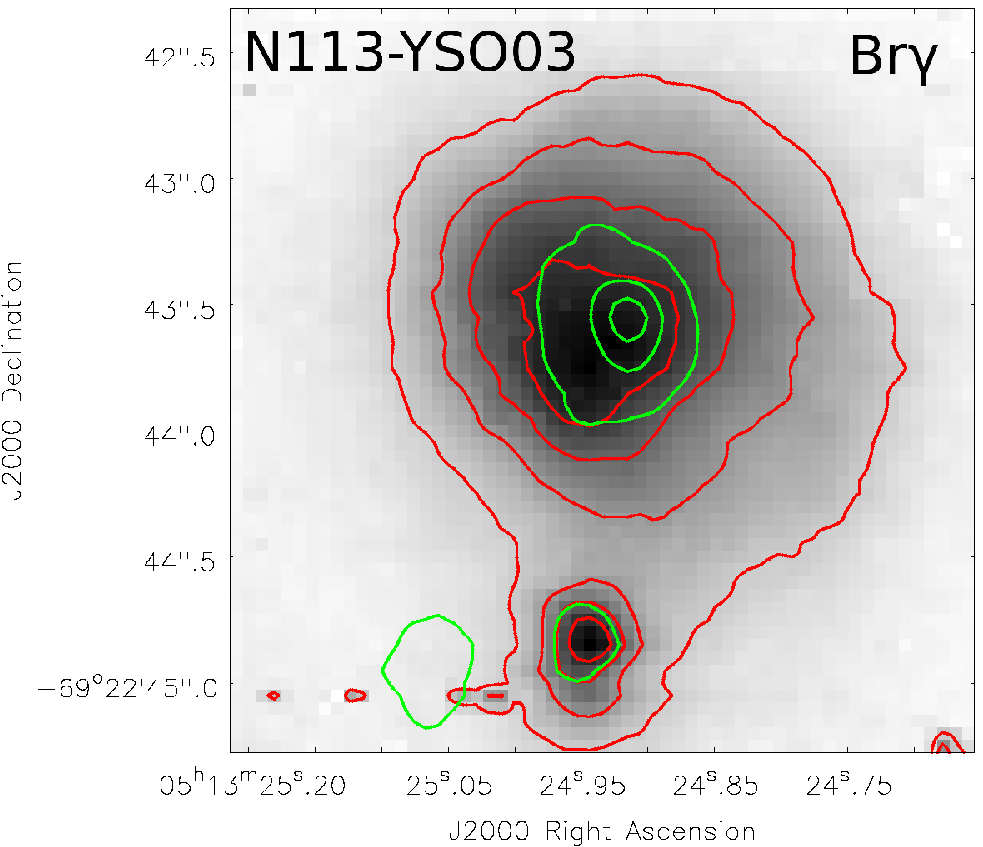}
  \includegraphics[width=0.32\linewidth]{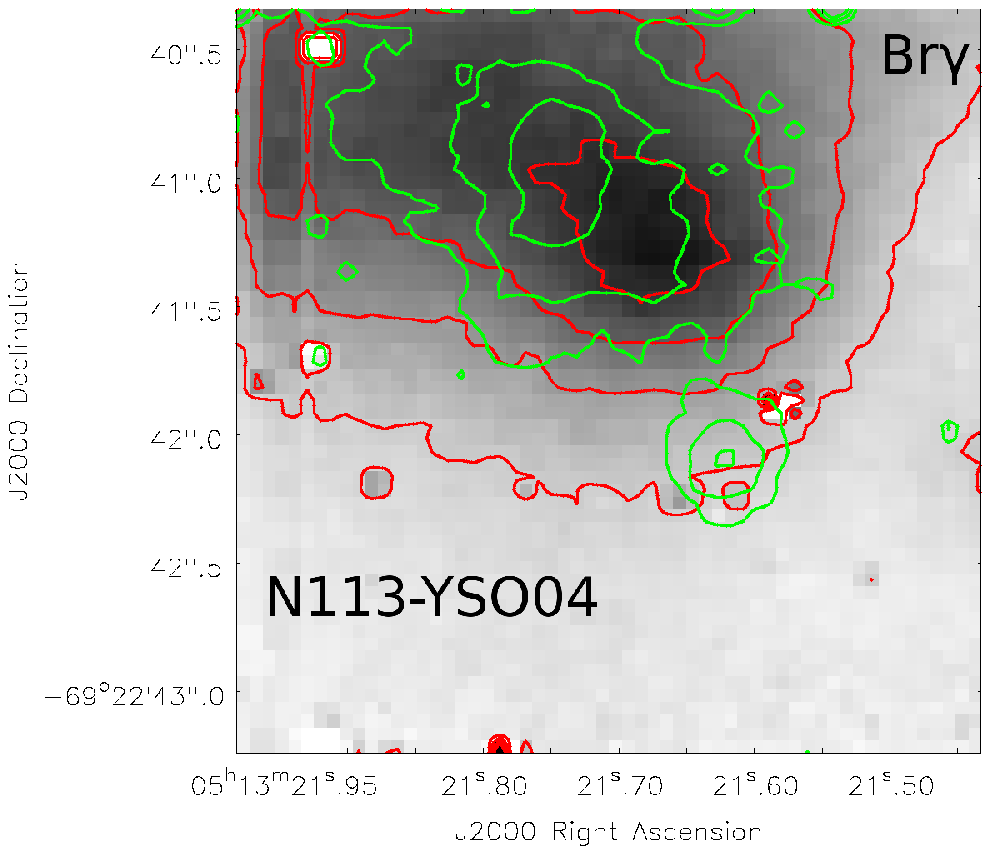} 
\includegraphics[width=0.32\linewidth]{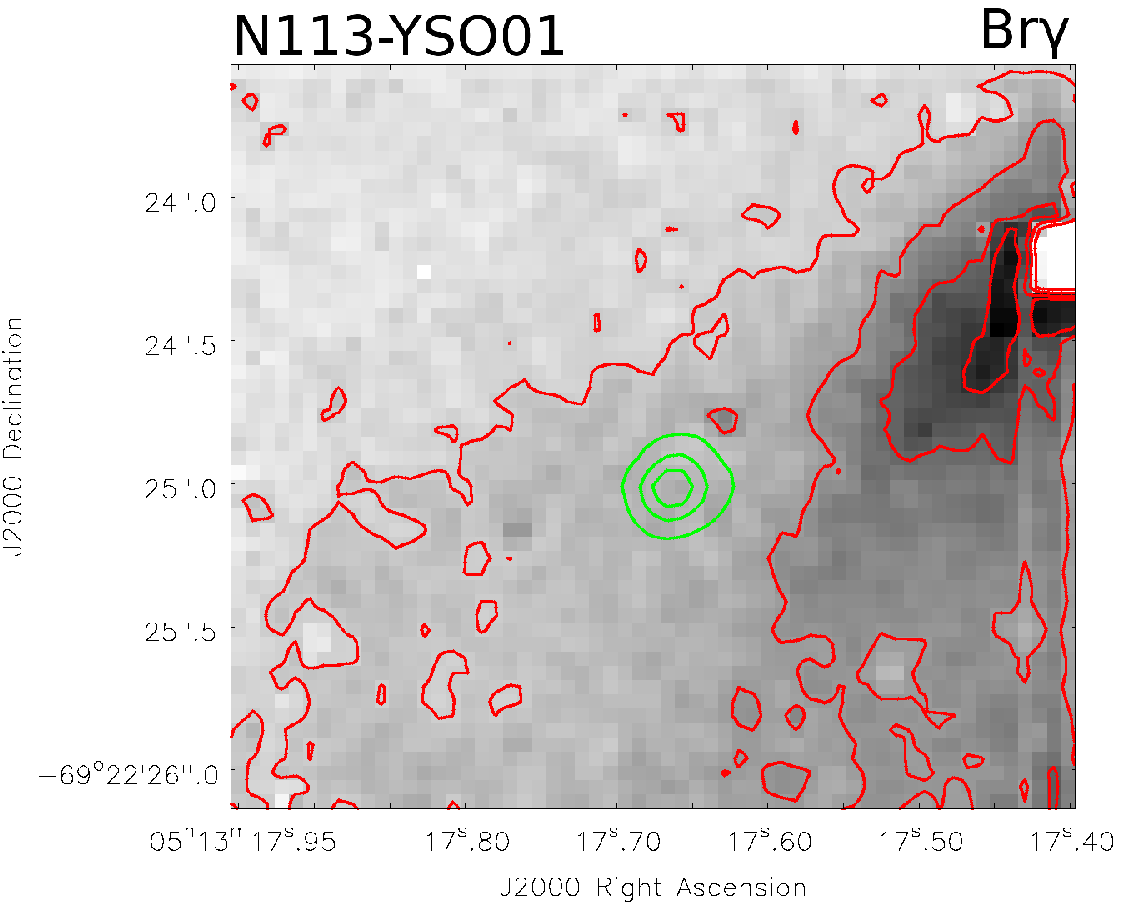} 
\includegraphics[width=0.32\linewidth]{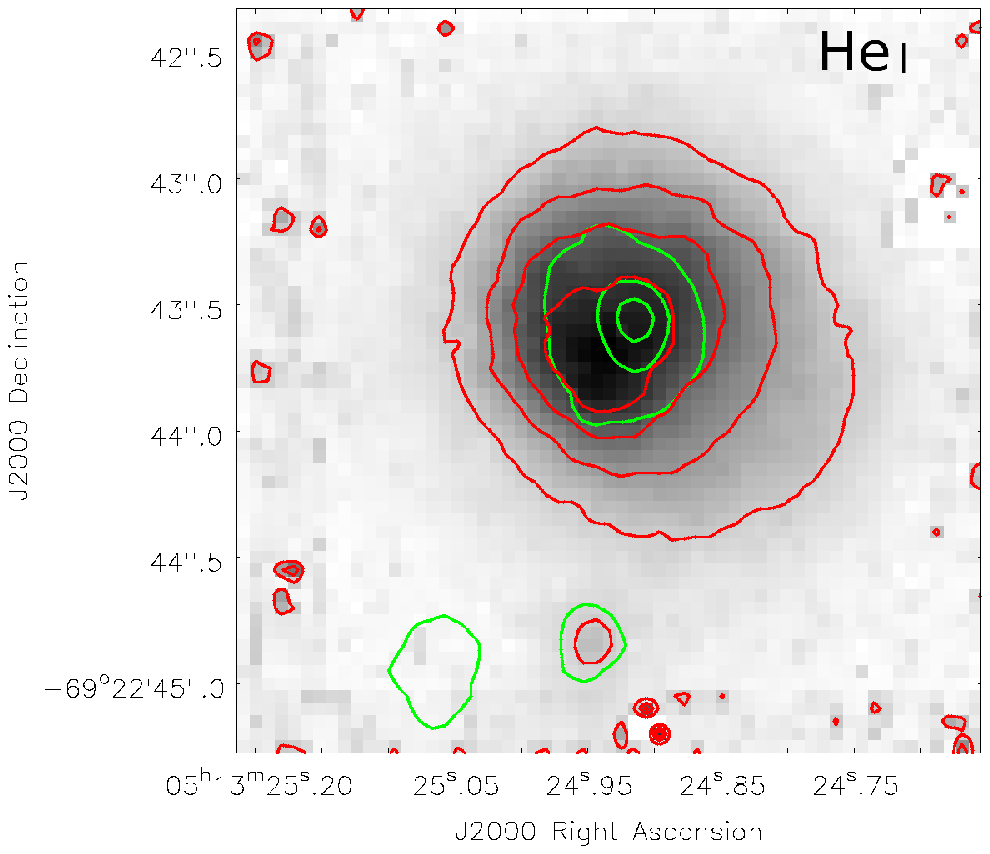}  
\includegraphics[width=0.32\linewidth]{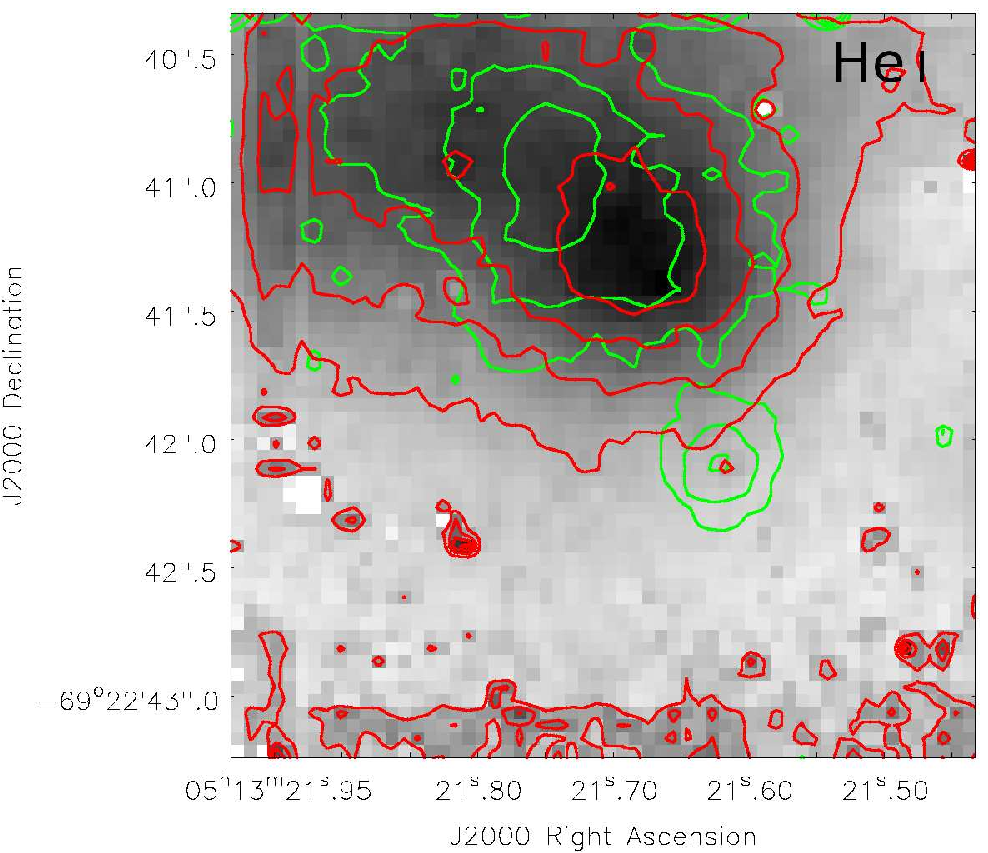} 
\includegraphics[width=0.32\linewidth]{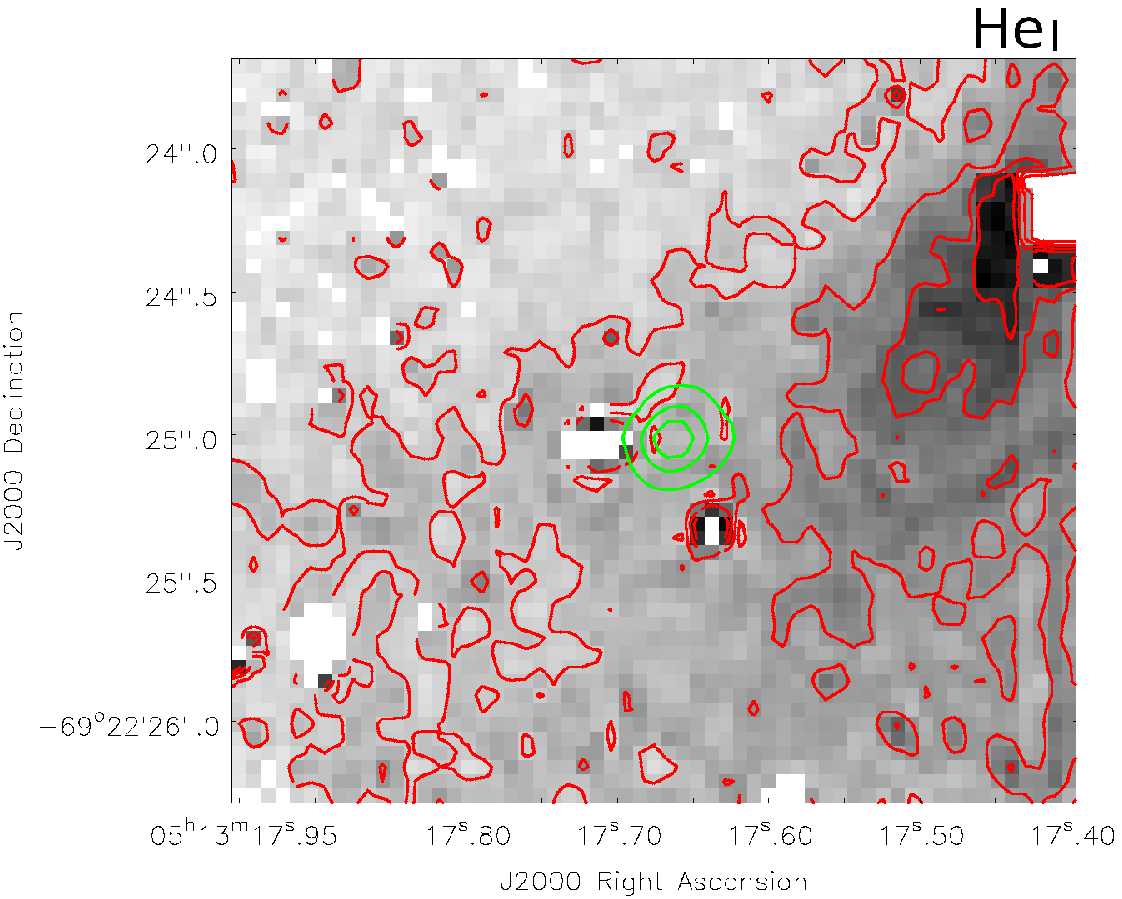}  
\includegraphics[width=0.32\linewidth]{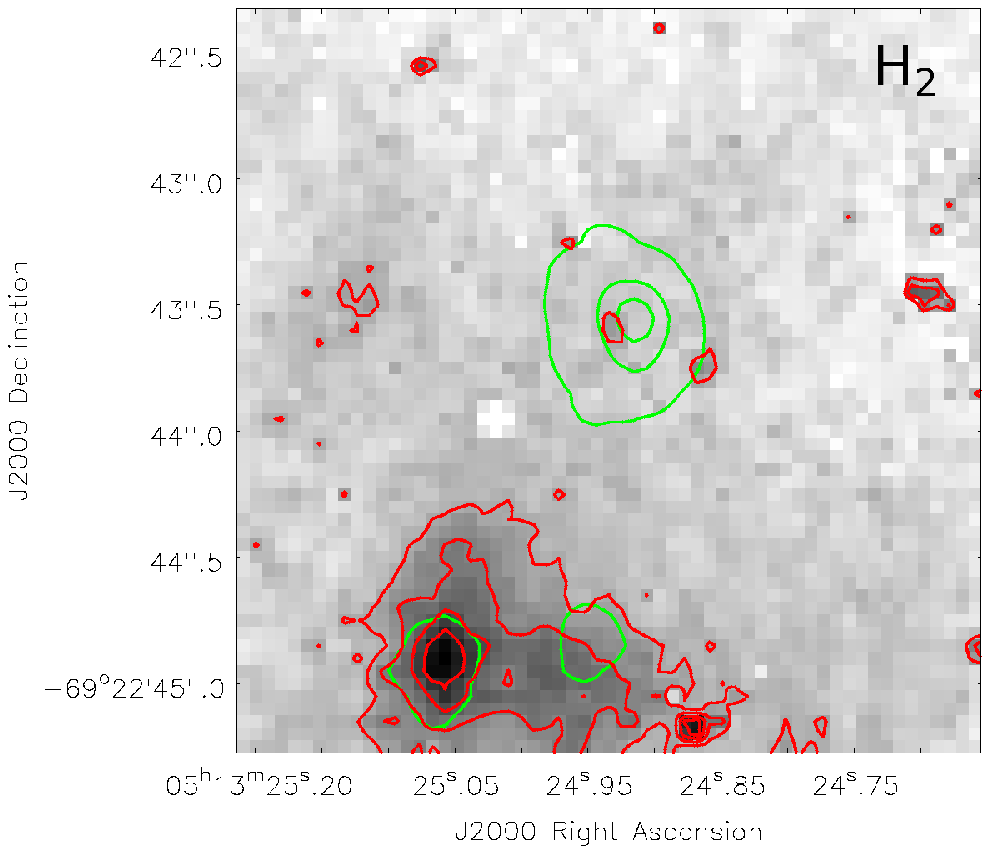}  
\includegraphics[width=0.32\linewidth]{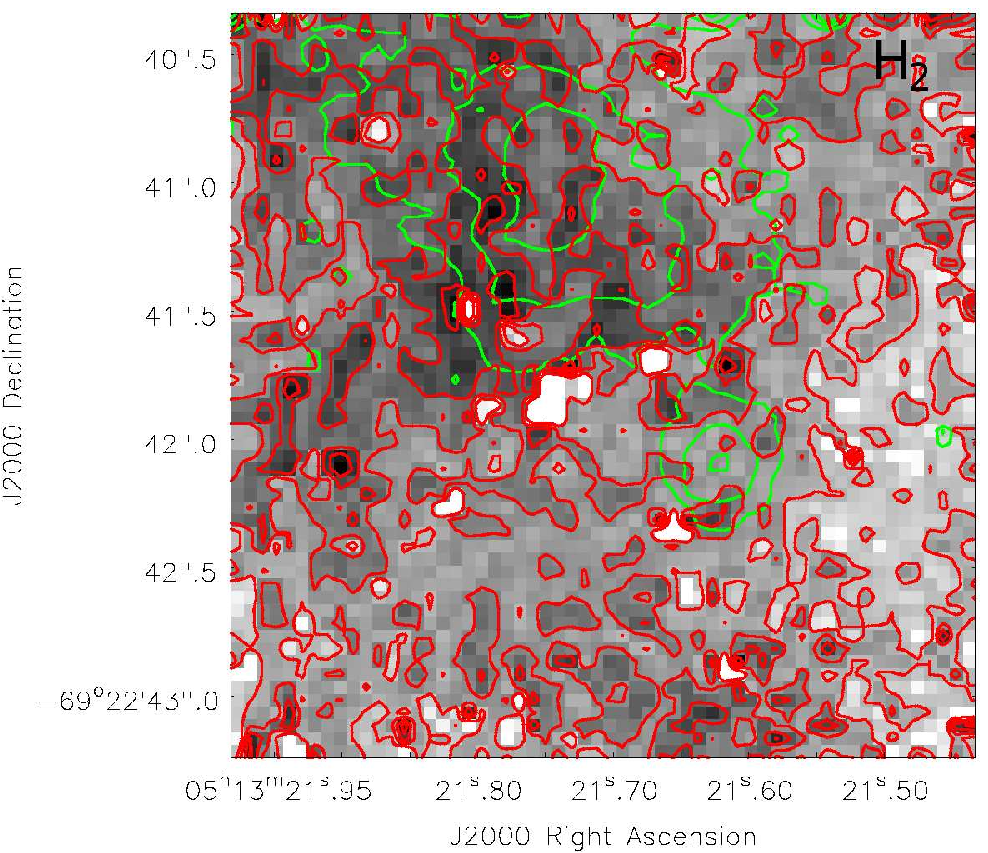}  
\includegraphics[width=0.32\linewidth]{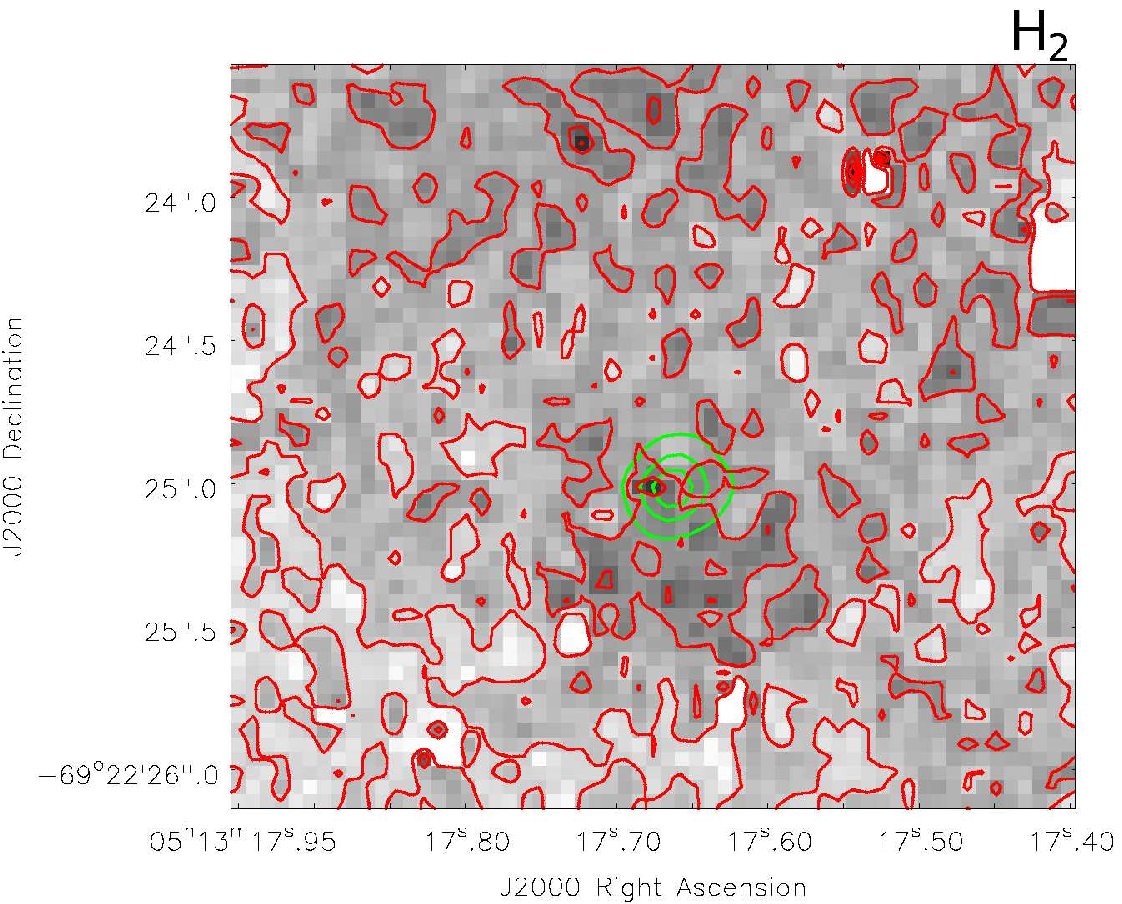}
\caption{Emission line maps with contours overlaid. Red contours - line emission [0.2, 0.4, 0.6, 0.8]$\times$peak, green contours - continuum emission 
[0.25, 0.5, 0.75]$\times$peak. Left to right - N113-YSO03, N113-YSO04, N113-YSO01. Top to bottom - Br$\gamma$, He {\sc i}, H$_2$ 1-0(S1).}
 \end{center}
 \end{minipage}
\end{figure*}

\subsubsection{H\,{\sc i} emission}

The strongest detected H\,{\sc i} emission line in this sample, Br$\gamma$, is most commonly 
associated with accretion in star formation studies. For intermediate mass YSOs the relation from \citet{Calvet2004}
can be used to estimate the accretion luminosity from Br$\gamma$ luminosity:
\begin{equation}
 \log(\text{L}_{\text{acc}}) = -0.7+0.9(\log(\text{L}_{\text{Br}\gamma})+4)
\end{equation}

Whilst this relation holds true for Herbig A stars, the higher mass Herbig B type stars have been observed to exhibit a Br$\gamma$ emission
excess \citep{Donehew2011, Mendigutia2011}, most likely due to an additional emission component originating from the strong 
winds driven by stellar UV photons emitted from hot stars. It is likely therefore that a similar effect is present in the more massive YSOs during the later phases
of their evolution. For the purposes of comparing our sample with a Galactic sample however, the above relation can be applied to gain an
equivalent accretion luminosity assuming that both samples cover the same range of evolutionary states and YSO masses.
Additionally we must consider the impact that metallicity may have on this relation. Whilst it is the case that the momentum and mass loss rates 
of stellar winds are strongly affected by metallicity \citep{Puls2000, Vink1999, Vink2001, Kudritzki2002, Krticka2006}, it is the number of photons 
produced which are able to ionise hydrogen that is significant when measuring Br$\gamma$ emission. \citet{Kudritzki2002} predicts that the 
number of photons capable of ionising hydrogen is barely affected by a change in metallicity. This suggests that the above relationship between 
Br$\gamma$ and accretion luminosity should hold for studies in lower metallicity environments.

Whilst bolometric luminosities have been obtained for each target using existing \textit{Spitzer} data (see Table 1), two out of three target fields contain
multiple continuum sources and the third exhibits emission that appears to originate from outside the FOV. Without higher resolution mid-infrared
studies it will not be possible to accurately determine bolometric luminosities for each of the sources. We therefore compare the equivalent accretion
luminosity with \textit{K}-band magnitude in Fig. 4 rather than bolometric luminosity. A distance of 49.4$\pm$0.5 kpc \citep{Laney2012}
was assumed to the LMC and distances to Galactic sources were obtained from the RMS survey database\footnote[4]{http://rms.leeds.ac.uk/cgi-bin/public/RMS\_DATABASE.cgi}.
From Fig. 4 we can ascertain that the Br$\gamma$ luminosities observed towards all six sources fall within 
the range observed in the Galactic sources. The sources for which we believe that the emission is associated with that source exhibit a higher 
Br$\gamma$ / \textit{K}-band continuum emission ratio than the remaining sources.
The accretion rates of the YSOs in this study are consistent with those of Galactic YSOs but appear to be high, possibly indicative of 
higher accretion rates.

\begin{figure*}
\begin{minipage}{175mm}
 \begin{center}
  \includegraphics[width=0.95\linewidth]{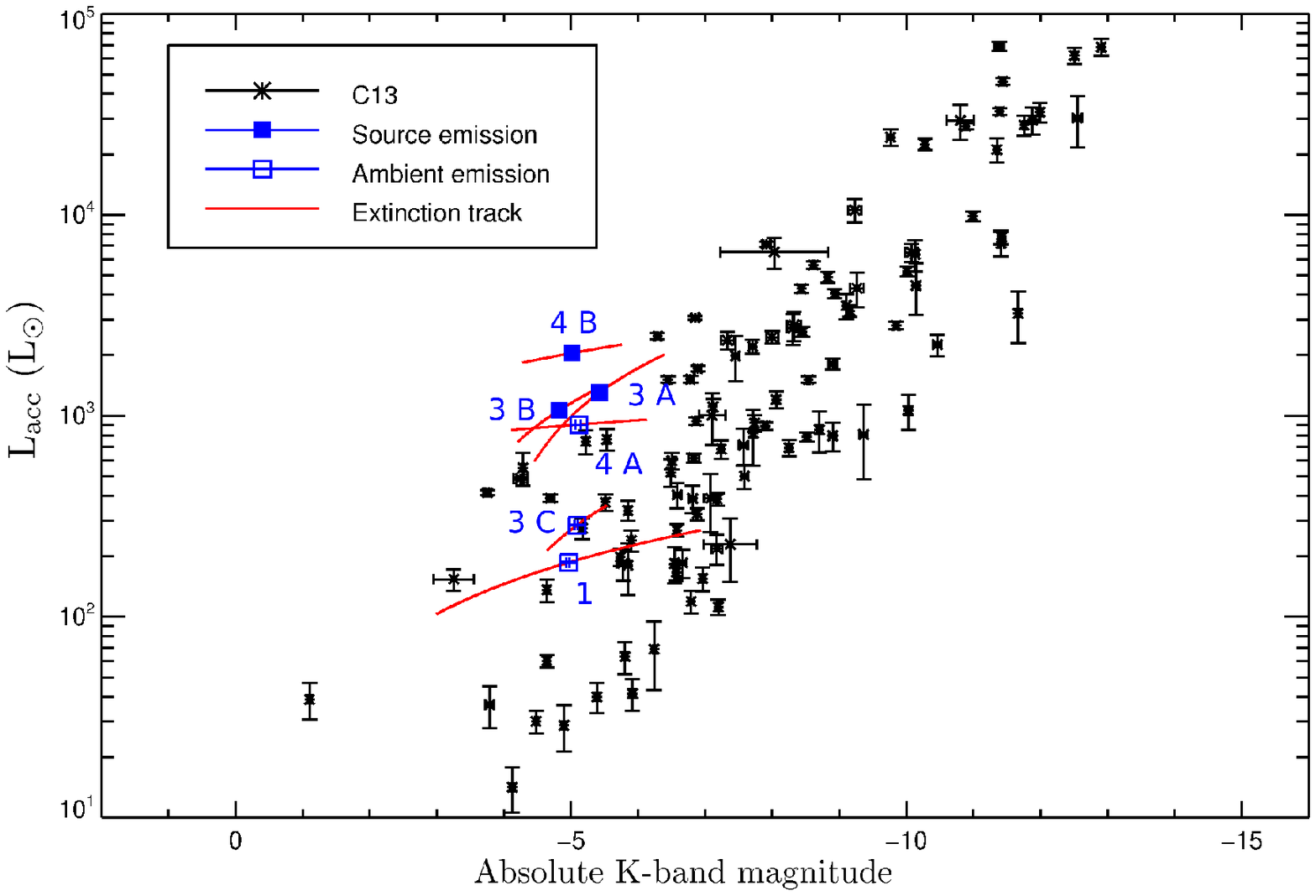}
\caption{
Equivalent accretion luminosity against absolute \textit{K}-band magnitudes. A distance to the LMC of 49.4$\pm$0.5 kpc is assumed.
 Extinction correction has been applied using $A_V$ values described in Section 4.2 for the N113 data and $A_V$ values from C13 for the Galactic data. Slit losses have not
been taken into account for the Galactic data. The range of possible values allowed by the uncertainty in extinction for each N113 source is shown as
a red extinction track. For clarity the N113-YSO prefixes have been omitted. Source emission (\textit{filled squares})
and ambient emission (\textit{open squares}) are discussed in the main text.}
 \end{center}
\end{minipage}
\end{figure*}

The spatial extent of the Br$\gamma$ emission is mapped in the top row of Fig. 3. Where the Br$\gamma$ emission is significantly spatially extended  
beyond the continuum source, it is likely that the contribution of non-accretion emission is significant. This appears to be the case in N113-YSO03 A
and N113-YSO04 B. The Br$\gamma$ emission is compact in N113-YSO03 B whilst in the remaining three continuum sources (N113-YSO01, N113-YSO03 C and N113-YSO04 A)
the Br$\gamma$ emission appears to be ambient or produced from other sources in the FOV.
The dominant source of Br$\gamma$ emission in N113-YSO01
 peaks outside of the FOV.

In addition to mapping the Br$\gamma$ emission flux around these sources, we have also mapped the centroid velocities relative to the centroid at the 
westernmost continuum source in each field (N113-YSO01, N113-YSO03 A, N113-YSO04 A), shown in Fig. 5. 
Spaxels where the uncertainty in relative velocity exceeds the imposed limits (5 km s$^{-1}$ in YSO03 and 10 km s$^{-1}$ in YSO01 and YSO04) have been masked.
Figure C1 shows emission line velocity maps obtained from the sky cube for N113-YSO03, showing that there are no significant systematic velocity gradients.
The two sources which exhibit extended Br$\gamma$ emission (N113-YSO03 A and N113-YSO04 B) exhibit clear velocity gradients ($\pm$10km s$^{-1}$ and $\pm$5km s$^{-1}$
respectively) centred on the continuum source,
suggesting the expansion of excited gas around these objects. The off-source Br$\gamma$ emission in the N113-YSO01 FOV appears to be slightly blueshifted
with respect to the central continuum source.

\begin{figure*}
 \begin{minipage}{175mm}
 \begin{center}
\includegraphics[width=0.495\linewidth]{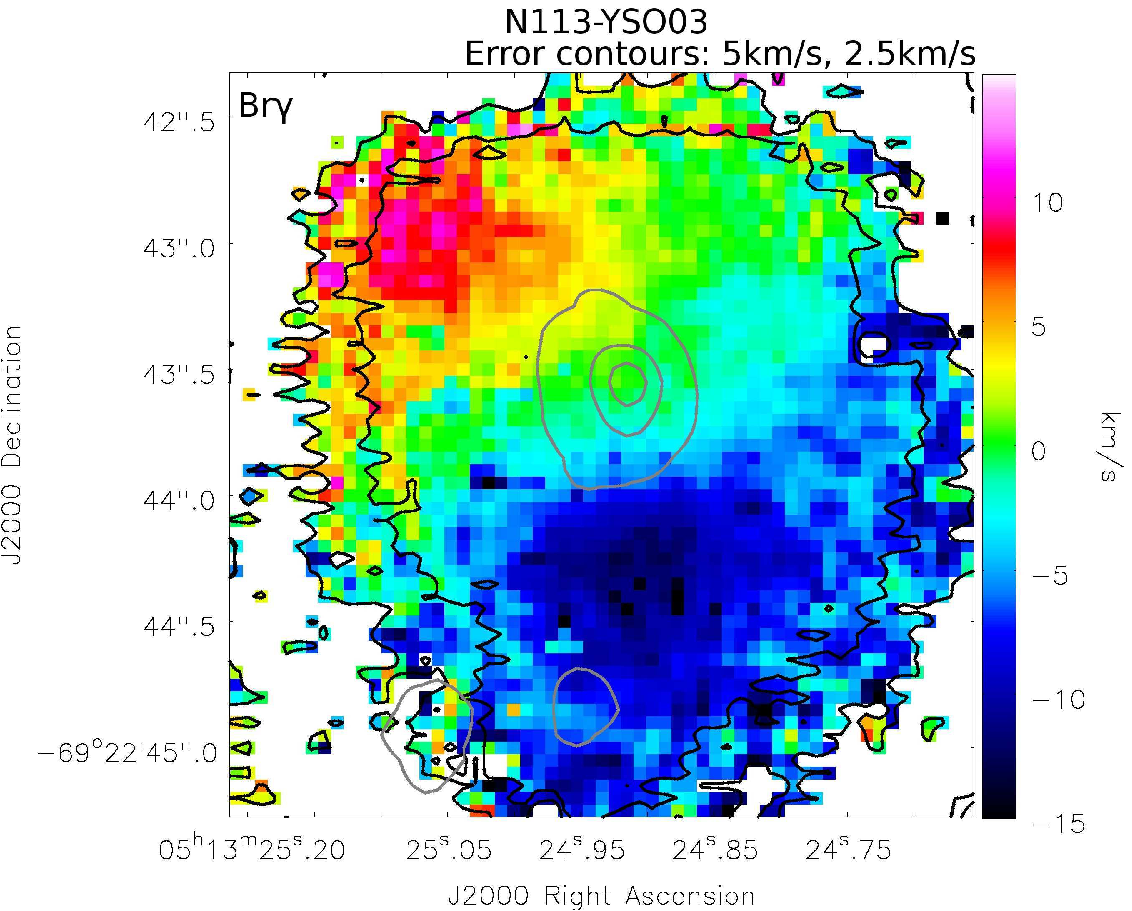}  
\includegraphics[width=0.495\linewidth]{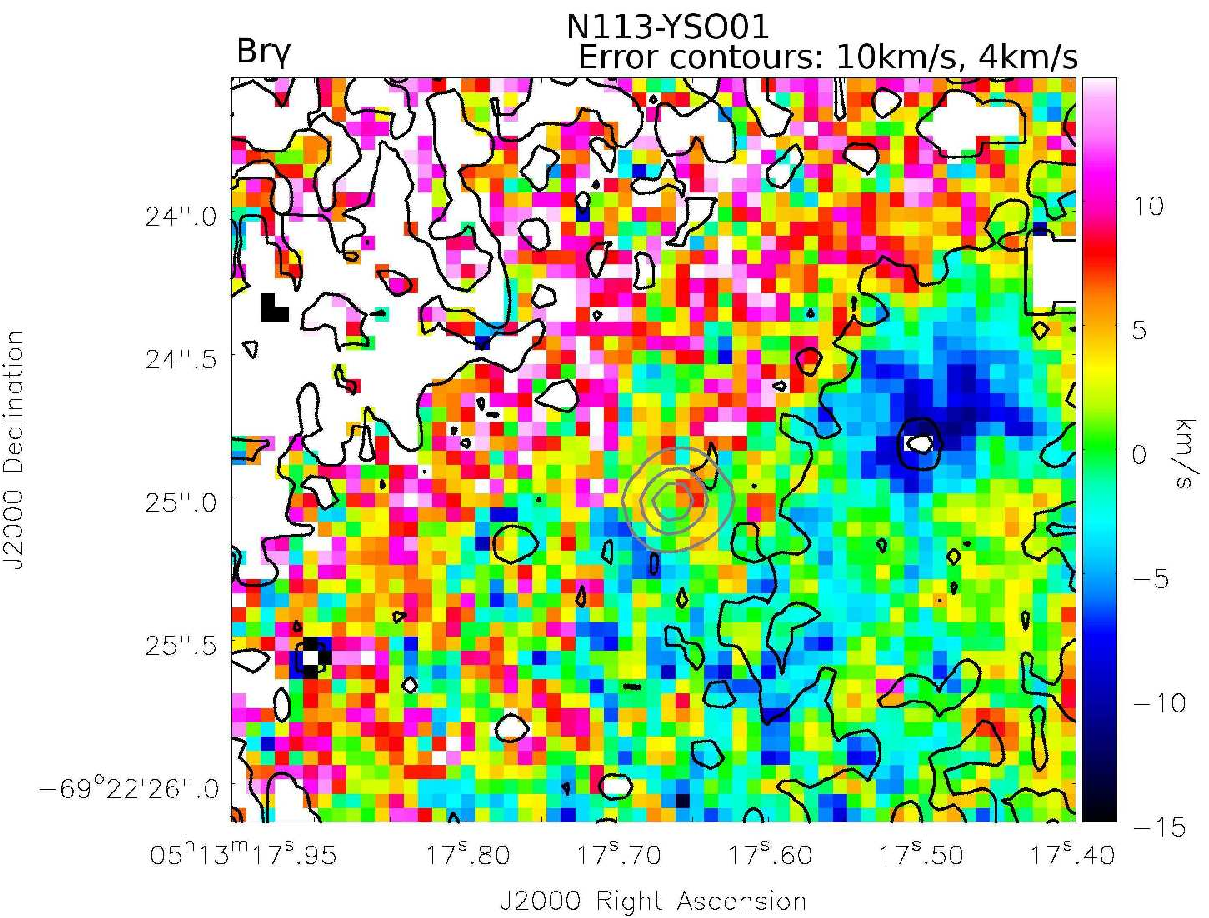}  
\includegraphics[width=0.495\linewidth]{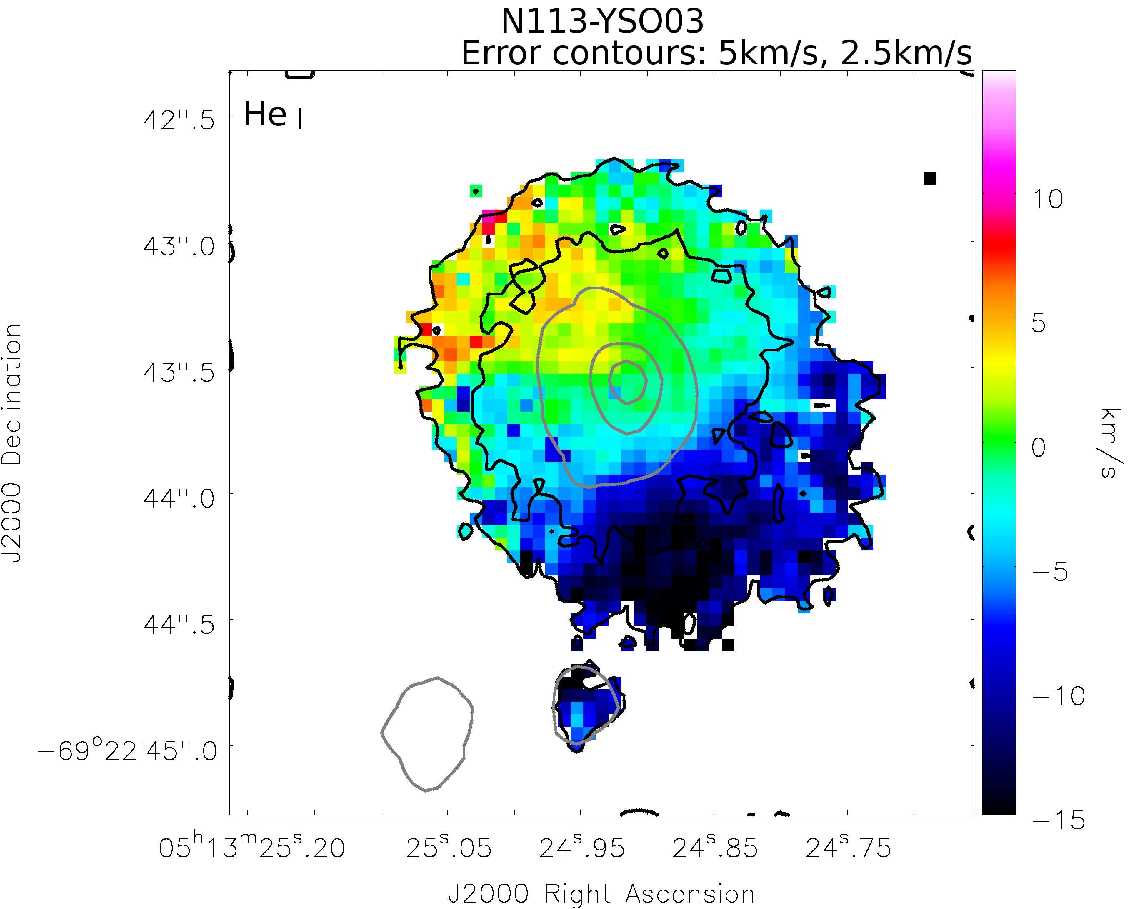}  
\includegraphics[width=0.495\linewidth]{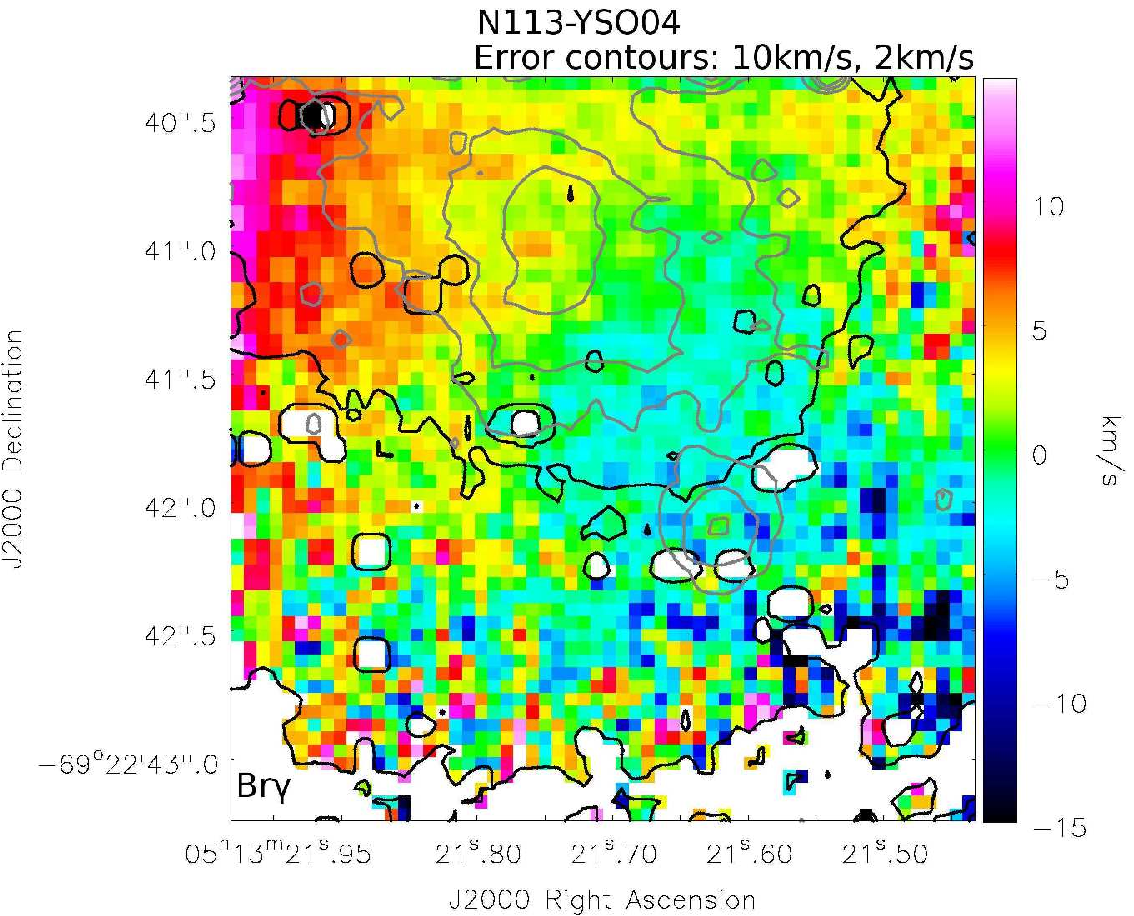} 
\includegraphics[width=0.495\linewidth]{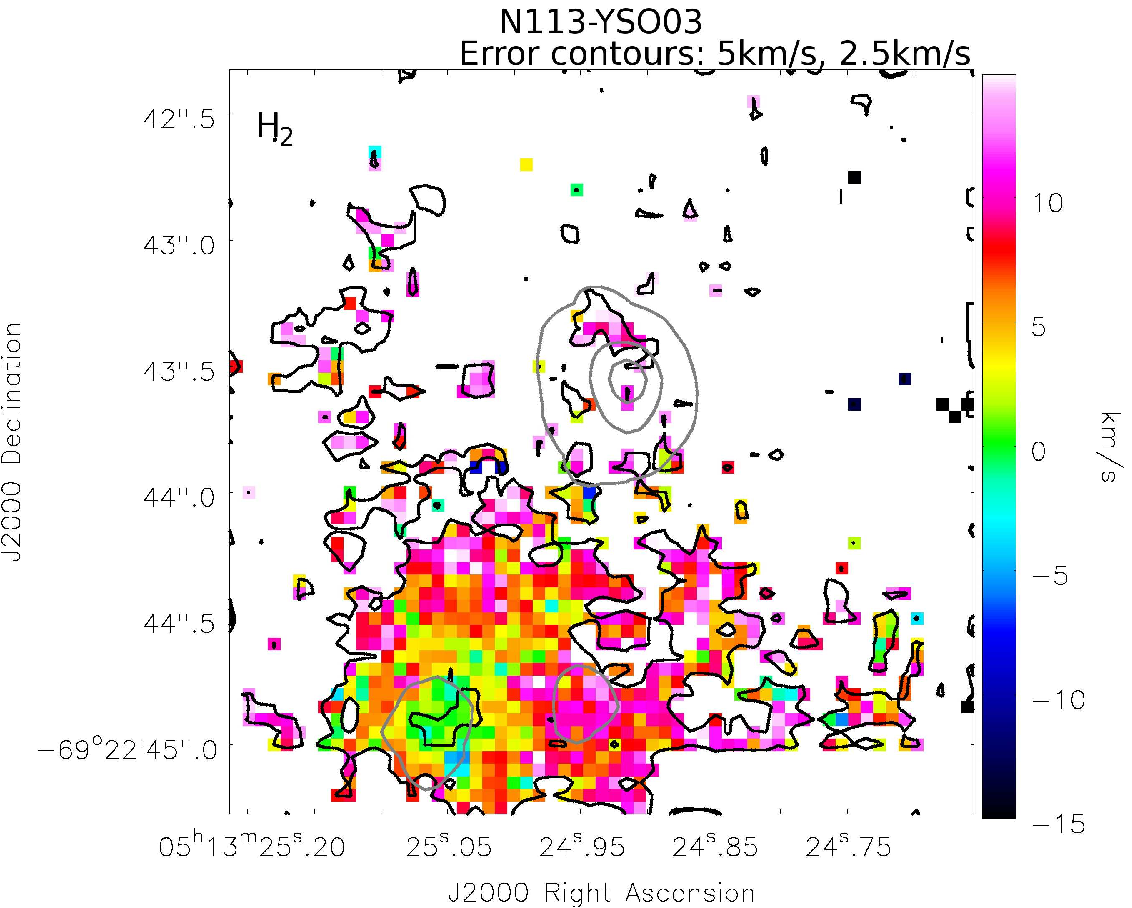} 
\includegraphics[width=0.495\linewidth]{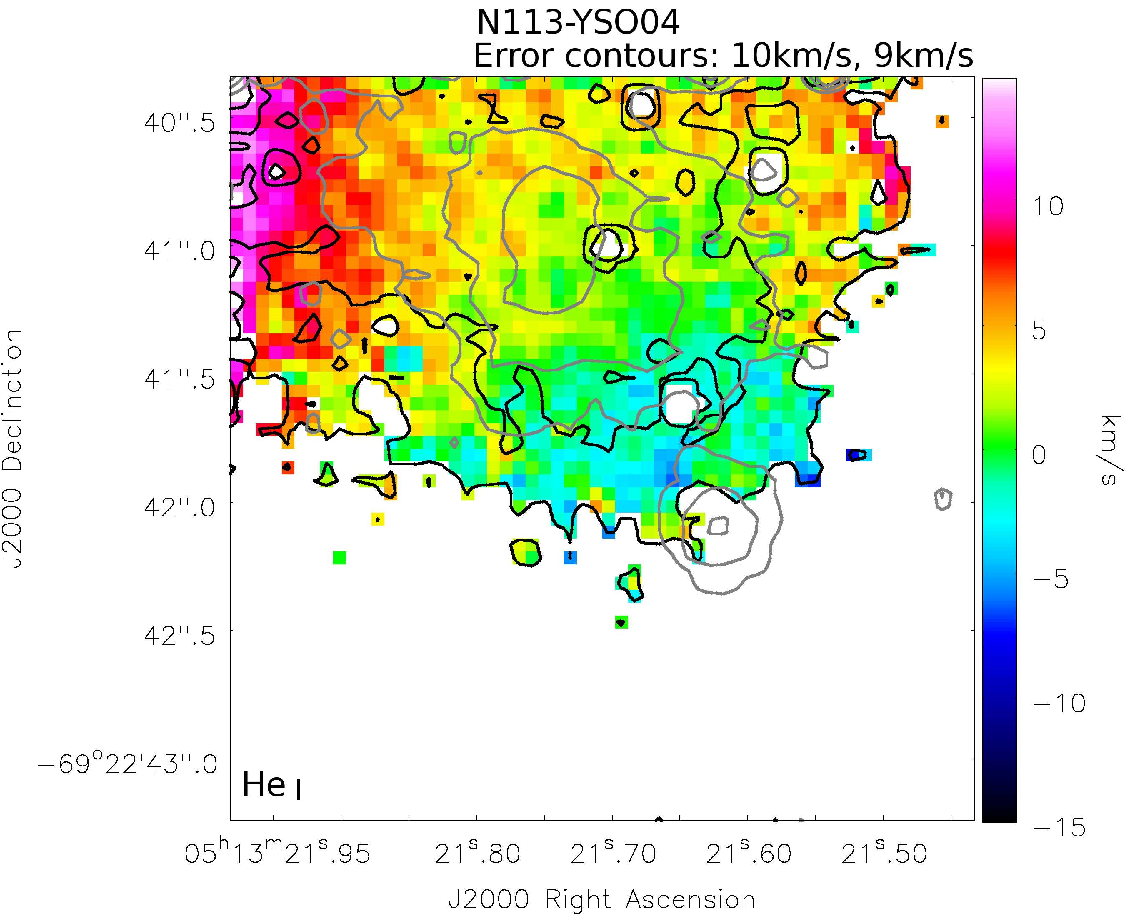}  

\caption{Left: Br$\gamma$, He {\sc i} and  H$_2$ 2.1218$\mu$m centroid velocity maps for N113-YSO03. 
Right: Br$\gamma$ centroid velocity map for N113-YSO01 (top) and Br$\gamma$ and He {\sc i} velocity maps for N113-YSO04 (middle and bottom).
Black contours represent the uncertainties; the outer (left) and inner (right) contour values are indicated in each image.
The continuum contour levels are [0.25, 0.5, 0.75]$\times$peak (grey).
}
 \end{center}
 \end{minipage}
\end{figure*}

The Pfund series is detected towards three of the six sources resolved in this work; N113-YSO03 A, N113-YSO03 B and N113-YSO04 B.
Unfortunately the \textit{K}-band Pfund series lies in an area of poor atmospheric transmission so it is not easy to obtain 
accurate measurements of their flux. The positions of the detected Pfund series emission lines are shown in Fig. A3 and 
the measured line fluxes are listed in Table B1.
Using the Pfund series emission and the Br$\gamma$ emission it is possible to obtain temperatures
from the ratios of hydrogen recombination lines if the density is well constrained. 
Whilst the measured line ratios are certainly consistent with the presence of massive OB-type stars, no further constraints on these 
physical parameters were obtained from our data, due the low S/N ratio in the Pfund series measurements, the large reddening 
uncertainties, and the strong dependence on density that is poorly constrained.

\subsubsection{He {\sc i} emission}

The primary production mechanism for the He {\sc i} emission line is the ionisation and subsequent 
recombination of helium which becomes significant at the ionisation boundary and potentially
in the collision with surrounding medium \citep{Porter1998}.  Whilst the He\,{\sc i} / Br$\gamma$ ratio is sensitive to the 
temperature of the emitting regions, its heavy dependance on density means that it cannot be used as a robust diagnostic
of temperature \citep{Shields1993}.

We detect the 2.0587 $\mu$m He {\sc i} emission line at the position of all six of the continuum sources and it is detected as extended 
emission around two sources (N113-YSO03 A and N113-YSO04 B; see Fig. 3, middle row). The two sources which exhibit the 
extended He {\sc i} emission are the strongest He {\sc i} emitters and those with the first and third
highest Br$\gamma$ fluxes, respectively. 
The He\,{\sc i} doublet at 2.113 $\mu$m was detected but not resolved towards three sources (N113-YSO03 A, N113-YSO04 A and N113-YSO04 B) indicating a
collisional excitation component in regions of high density \citep{Lumsden2001}. 
The flux of the 2.113$\mu$m doublet is typically significantly lower than the 2.058 $\mu$m line (Table B1) and the S/N for the doublet in our sample prevents further
analysis. No He\,{\sc ii} emission has been detected towards any of the sources in this work.

\begin{figure*}
\begin{minipage}{175mm}
 \begin{center}
  \includegraphics[width=0.8\linewidth]{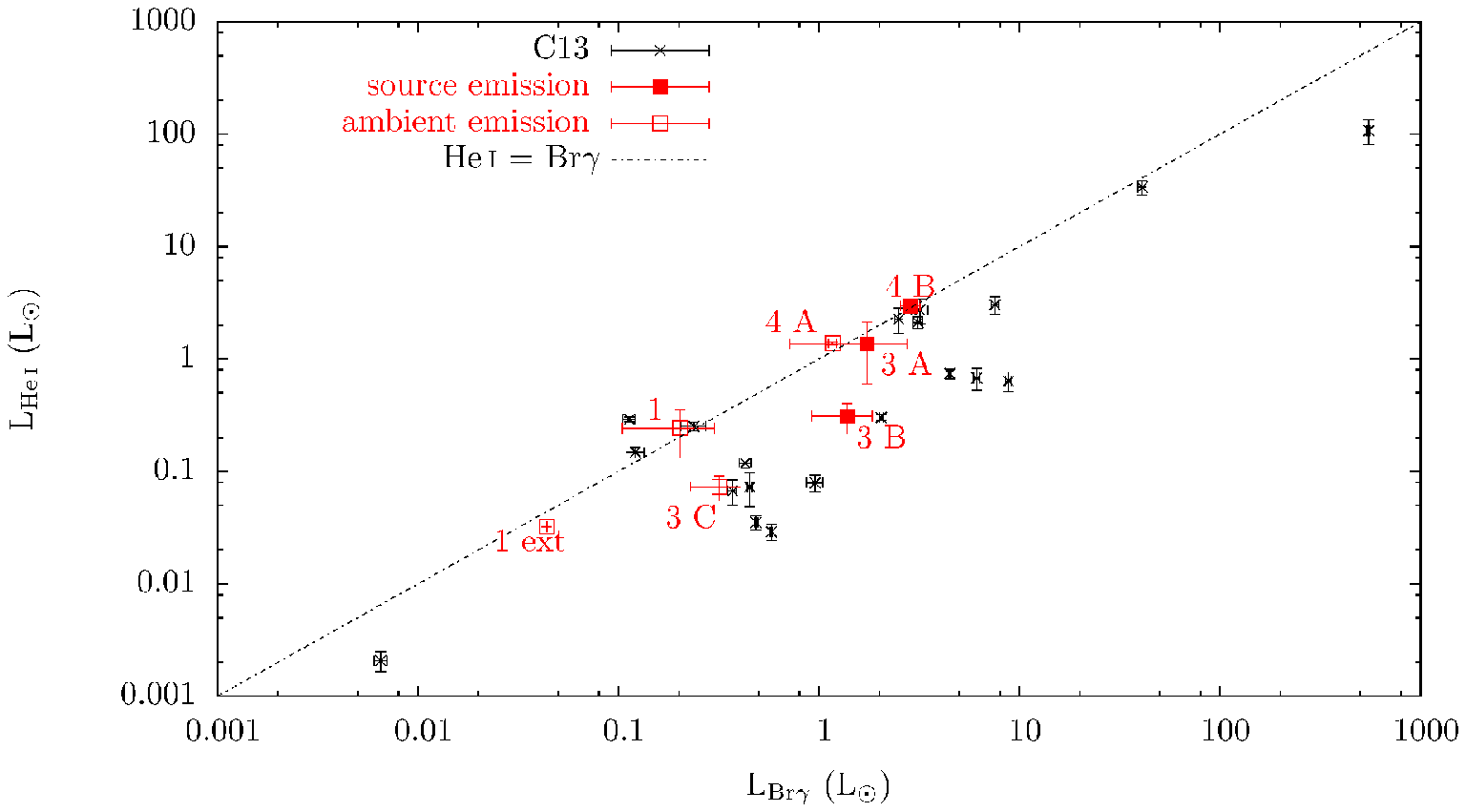}
 \end{center}
\caption{He {\sc i} luminosity plotted against Br$\gamma$ luminosity for all continuum sources. 1 ext denotes the extended emission in the N113-YSO01
cube (not corrected for extinction). For clarity the N113-YSO prefixes have been omitted.}
\end{minipage}
\end{figure*}

Spatially, the He {\sc i} emission tends to trace the same structures as the Br$\gamma$ emission, although with weaker and slightly more compact emission.
This is to be expected in sources where the central source is hot enough to excite a large volume of surrounding gas as in a compact
H\,{\sc ii} region.
In N113-YSO03 C and N113-YSO04 A, the detected He {\sc i} emission appears to be ambient to the region, possibly originating from N113-YSO03 A and N113-YSO04 B, respectively.
The He\,{\sc i} emission in N113-YSO01 traces the morphology of the Br$\gamma$ emission, also appearing to originate outside of the FOV.

Figure 6 plots the He {\sc i} 2.0587 $\mu$m emission line luminosity against the Br$\gamma$ luminosity for all N113 continuum sources and those from C13 for 
which both line measurements are available. It appears that the same trend and range of values is observed in N113 as in the Milky Way.
The off source emission in N113-YSO01 (indicated by \textquotedblleft1 ext\textquotedblright) appears to have a comparable He {\sc i}/Br$\gamma$ ratio to N113-YSO03 A
and N113-YSO04 B, suggesting that it presents a relatively energetic environment.

As well as tracing the same morphological structures, where extended He {\sc i} emission is present it exhibits the same velocity fields as the Br$\gamma$ 
emission, as demonstrated in Fig. 5 for N113-YSO03 and N113-YSO04. This is further evidence that the extended Br$\gamma$ emission and extended He {\sc i}
emission originate from the same strong radiation field.

\subsubsection{H$_2$ emission}

The H$_2$ lines detected in our spectra are identified in Fig. A1.
Although the H$_2$ emission has been spatially mapped (see Fig. 3), the signal-to-noise per spaxel in N113-YSO01 and N113-YSO04 for the H$_2$ lines is poor
and very little morphological information can be obtained. 
Additionally it appears likely that the H$_2$ emission in N113-YSO01 and N113-YSO04 is consistent with uniform ambient H$_2$ emission, unrelated to the
discrete YSOs.
In N113-YSO03 the H$_2$ emission is relatively compact and peaks at the position of source C.
The H$_2$ emission measured in source B may have a significant component originating from source C. Whilst N113-YSO03 A does not show
significant H$_2$ emission in Fig. 3, on inspection of the extracted spectrum it does exhibit relatively weak H$_2$ emission which is likely to be ambient,
as is the case for N113-YSO01 and N113-YSO04.
In summary only N113-YSO03 C is a significant source of H$_2$ emission.

The H$_2$ 2.1218 $\mu$m emission luminosity is plotted against the Br$\gamma$ luminosity for each of the continuum sources in Fig. 7. Little correlation can
be seen on this diagram, suggesting that the emitting regions of the lines are unrelated. N113-YSO03 C clearly falls above the H$_2 = $ Br$\gamma$ line 
whereas Br$\gamma > $ H$_2$ for all of the other sources, indicating that N113-YSO03 C is dominated by H$_2$ emission whilst the 
remaining sources are dominated by atomic Br$\gamma$ emission, consistent with the view that the H$_2$ 
emission in N113-YSO01, N113-YSO04 and possibly N113-YSO03 A is mostly ambient.

\begin{figure*}
\begin{minipage}{175mm}
 \begin{center}
\includegraphics[width=0.8\linewidth]{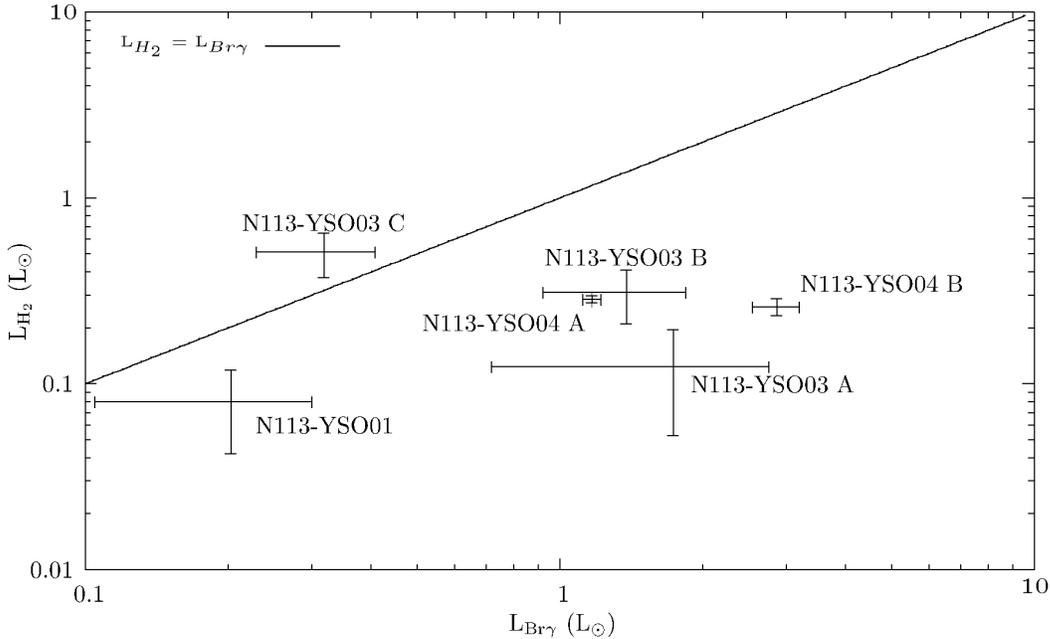}
\caption{H$_2$ 2.1218$\mu$m emission against Br$\gamma$ emission for all observed continuum sources.}
 \end{center}
\end{minipage}
\end{figure*}

\begin{figure*}
\begin{minipage}{175mm}
 \begin{center}
\includegraphics[width=0.85\linewidth]{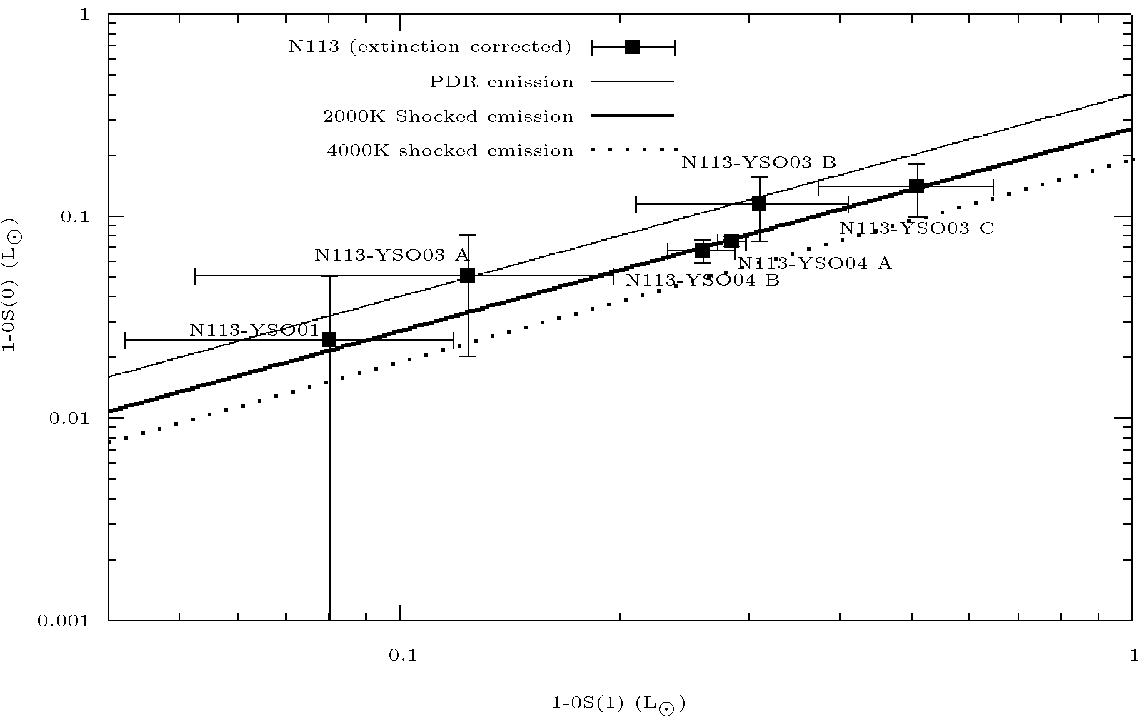}
\includegraphics[width=0.85\linewidth]{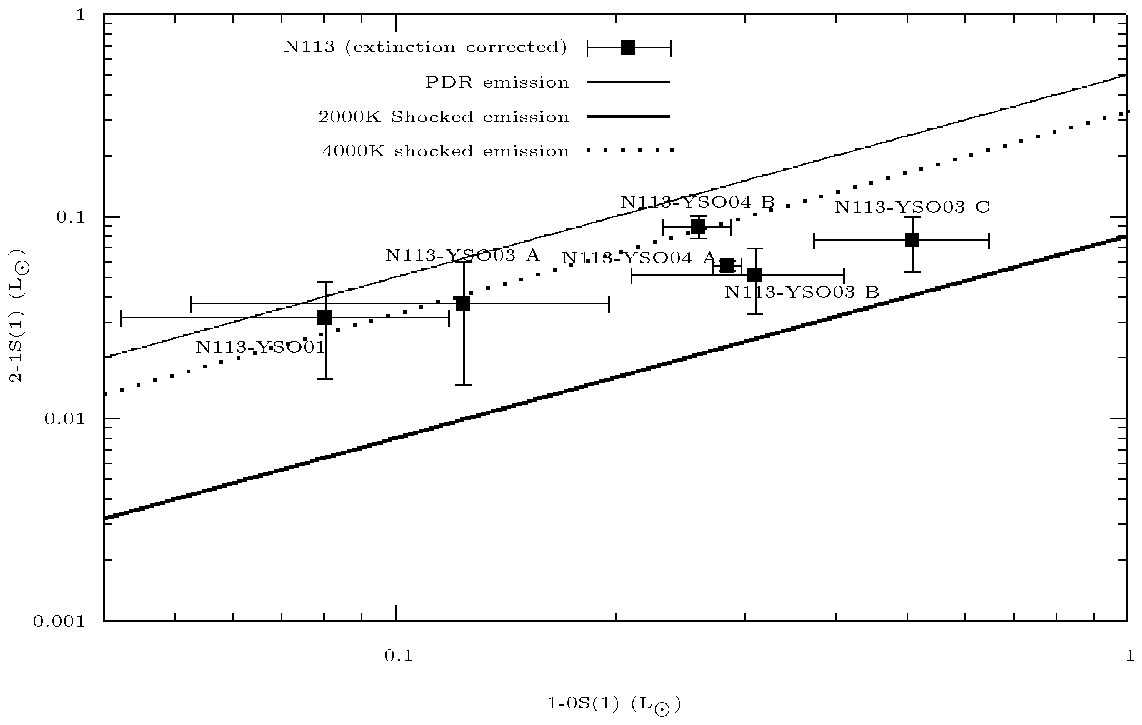}
\caption{Top: H$_2$ 1-0(S0) 2.2235 $\mu$m emission against H$_2$ 1-0(S1) 2.1218 $\mu$m emission for all observed continuum sources.
Bottom: H$_2$ 2-1(S1) 2.2477 $\mu$m emission against H$_2$ 1-0(S1) 2.1218 $\mu$m emission for all observed continuum sources. Lines shown are for 
photodissociation emission and shocked emission at 2000 K and 4000 K as indicated in the legend.}
 \end{center}
\end{minipage}
\end{figure*}

\begin{table*}
\begin{minipage}{175mm}
 \begin{center}
\caption{Extinction corrected H$_2$ emission line ratios with respect to the 1-0S(1) emission line for all objects. Also included are the expected 
H$_2$ line ratios for photodissociation \citep{Black1987} and shocked emission \citep{Shull1978}.}
  \begin{tabular}{l l l l l l l}
\hline
   Object & 1-0S(0) & 1-0S(2) & 1-0S(3) & 2-1S(1) & 2-1S(2) & 2-1S(3)\\
  & 2.2235 $\mu$m & 2.0338 $\mu$m & 1.9576 $\mu$m & 2.2477 $\mu$m & 2.1542 $\mu$m & 2.0735 $\mu$m \\
\hline
N113-YSO01 & 0.31$\pm$0.35 & 0.68$\pm$0.57 & \null & 0.39$\pm$0.27 & 0.18$\pm$0.12 &\null \\
N113-YSO03 A & 0.41$\pm$0.34 & 0.36$\pm$0.28 & 0.44$\pm$0.35 & 0.30$\pm$0.25 & 0.15$\pm$0.12 & 0.34$\pm$0.27\\
N113-YSO03 B & 0.37$\pm$0.18 & 0.30$\pm$0.13 & 0.24$\pm$0.10 & 0.17$\pm$0.08 & 0.06$\pm$0.03 & 0.21$\pm$0.09\\
N113-YSO03 C & 0.28$\pm$0.11 & 0.34$\pm$0.12 & 0.45$\pm$0.16 & 0.15$\pm$0.06 & 0.06$\pm$0.02 & 0.17$\pm$0.06\\
N113-YSO04 A & 0.26$\pm$0.02 & 0.46$\pm$0.04 & \null & 0.20$\pm$0.02 & 0.15$\pm$0.02 & 0.23$\pm$0.02\\
N113-YSO04 B & 0.26$\pm$0.04 & 0.53$\pm$0.09 & \null & 0.34$\pm$0.06 & 0.19$\pm$0.03 & 0.40$\pm$0.06\\
\hline
Photoexcitation & 0.4--0.7 & 0.4--0.6 & & 0.5--0.6 & 0.2--0.4 & 0.2--0.3 \\
1000 K shock &  0.27 & 0.27 &  0.51 & 0.005 & 0.001 & 0.003 \\
2000 K shock & 0.21 & 0.37 & 1.02 &  0.083 & 0.031 & 0.084 \\
3000 K shock & 0.19 & 0.42 & 1.29 & 0.21 & 0.086 & 0.27 \\
4000 K shock & 0.19 &  0.44 & 1.45 & 0.33 & 0.14 & 0.47 \\
\hline
  \end{tabular}

 \end{center}
\end{minipage}
\end{table*}

The ratios of \textit{K}-band H$_2$ line fluxes can be used to determine whether the source of the emission is photodissociation or shock excited.
The extinction corrected flux ratios for all of the observed H$_2$ lines with respect to the 1-0(S1) line are shown in Table
4, with the expected ratios based on models of photoexcited emission and interstellar shocks in molecular clouds at various temperatures given in the lower part of 
Table 4. Whilst it is likely that all targets have contributions
from both shocked and a Photo-Dissociation Region (PDR), we can use these values to determine which excitation mechanism of H$_2$ emission is dominant.
 In the radiative excitation scenario, the ratios with respect to the 2.1218 $\mu$m 1-0(S1) line should fall in the range 0.5 to 0.6 and 0.4 to 0.7 for 
the 2-1(S1) and 1-0(S0) lines (at 2.2477 $\mu$m and 2.2235 $\mu$m), respectively \citep{Black1987}. Adopting $T = 2000$ K, shock-excitation should give values 
of 0.08 and 0.21, respectively for the same emission line ratios \citep{Shull1978}. 
Together, these two line ratios form a powerful diagnostic tool as the 2-1(S1) line ratio has a significant shock temperature dependance
whereas the 1-0(S0) ratio exhibits a large gap between predicted PDRs and shocks and, crucially, an inverse dependance on shock temperature.

In Fig. 8 we compare our observed 1-0(S0) and 2-1(S1) ratios relative to the 1-0(S1) line to those expected from photoexcitation, and 
shocked emission at 2000 K and 4000 K.
We find that in the upper panel three sources appear highly likely to be shock emission dominated (N113-YSO03 C, N113-YSO04 A, N113-YSO04 B) and two sources 
could be PDR emission dominated (N113-YSO03 A, N113-YSO03 B) but the uncertainties are large and therefore they could also fall within the shocked regime.
 For the final source (N113-YSO01), the uncertainty in the 1-0(S0) flux is such that no conclusions can be drawn on the origin of the emission. 
In the lower panel of Fig. 8, all of the sources appear to fall within the shocked emission zone, with two of the sources (N113-YSO01, N113-YSO03 A) within 
1$\sigma$ of PDR emission. 
From this we can draw the conclusion that for N113-YSO03C (the only source for which H2 emission is not ambient) the emission is shock-dominated.
The H$_2$ emission from sources A and B in N113-YSO03 appears to have contribution from both shock excitation and photodissociation whilst towards
both sources in N113-YSO04 the emission appears to be shock dominated. Due to low S/N no conclusion can be drawn for the source of the H$_2$ emission
in N113-YSO01.

N113-YSO03 was the only source for which we were able to make H$_2$ centroid velocity measurements (Fig. 5) and over the relatively small spatial
range where this was possible, there appears to be a small velocity gradient towards the red in the emission West of source C.

\section{Discussion}

\subsection{N113-YSO01}

The N113-YSO01 FOV contains a single VLT \textit{K}-band continuum source to a resolution of $0\rlap{.}^{\prime\prime}2$. \textit{Spitzer} resolution at the 
shortest wavelengths is $\sim$2$^{\prime\prime} $ so whilst
the bolometric luminosity ($1.51\substack{+0.78 \\ -0.25}$ $\times$10$^{5}$ L$_{\odot}$) for this source is likely to be more robust, 
the emission lines from the spatially unresolved studies will likely be contaminated by the off source emission seen in Fig. 3.
Both the Br$\gamma$ emission and the He\,{\sc i} emission peak approximately $1\rlap{.}^{\prime\prime}3$ ($\sim$0.31 pc) away from the continuum source. 
It appears that all of the measured line emission towards this object is ambient and if line emission from the source is present it is too weak to be
detected in this work.
In the top panel of Fig. 1, two large H$\alpha$ bubbles can be seen in close proximity to N113-YSO01. Thus it is possible 
that the line emission detected towards N113-YSO01 is associated with these larger scale structures.

Figure 1 shows the positions of three maser sources that are close to N113-YSO01, however
spectroscopically N113-YSO01 does not exhibit any indicators of strong outflows which could stimulate the maser emission.
Additionally the distances (0.52 pc -- 1.01 pc)
between these masers and the N113-YSO01 infrared source suggests that the masers may be in fact associated with 
other, weaker \textit{K}-band sources just visible in Fig. 1.
The maser sources also do not appear to coincide
with the prominent \textit{Spitzer} source. 

Massive YSOs without any significant \textit{K}-band emission lines are not common in the Galactic data set but they have been observed (e.g. G103.8744+01.8558, C13).
Longer wavelength data available for those Galactic sources makes the YSO identification credible.
It is possible that such sources represent the youngest hot core objects or they are simply weakly emitting YSOs.
However, the possiblity that the source observed in the N113-YSO01 FOV is not in fact a YSO cannot be excluded by this work. At this stage we are unable to 
constrain the origin of the line emission in the \textit{K}-band and \textit{Spitzer}-IRS spectra.

\subsection{N113-YSO03}

The line morphologies observed in the N113-YSO03 FOV suggest that the three continuum sources represent three different stages of star formation, 
likely reflecting a range of masses.
N113-YSO03 A is the brightest of all the observed sources in the \textit{K}-band with strong, extended and expanding Br$\gamma$ emission and He {\sc i} 
emission, implying that it is a relatively evolved compact H\,{\sc ii} 
region.
It also has the strongest He\,{\sc i} emission, likely caused by the strong winds associated with an emerging massive star.
 
N113-YSO03 B also exhibits strong Br$\gamma$ emission but it is compact and the He\,{\sc i} emission is much weaker, suggesting a considerably
lower contribution to the Br$\gamma$ emission from an H\,{\sc ii} region and thus a cooler and likely earlier stage (younger or less massive) object than N113-YSO03 A, possibly dominated by
accretion emission. 

N113-YSO03 C exhibits very weak or absent Br$\gamma$
and He {\sc i} emission, and  very strong H$_2$ emission. In addition, the H$_2$ line ratios are consistent with emission that is very heavily
dominated by shocked emission with temperatures in the range 2000--3000 K. It is therefore likely that N113-YSO03 C is the least evolved of the three sources, representing
an early stage YSO with strong bipolar outflows. 
This is also consistent with the much redder continuum slope for its spectrum.
The highest intensity water maser source in the Magellanic Clouds was identified 
at a distance of 0.36$\pm$0.07 pc (1$\rlap{.}^{\prime\prime}$5$\pm$0$\rlap{.}^{\prime\prime}$3) of source N113-YSO03 C \citep{Lazendic2002, Oliveira2006, Carlson2012}.
Of the three sources resolved in this FOV, N113-YSO03 A is the most massive, having evolved to the point of shaping its environment in the form of a
 compact H\,{\sc ii} region, while N113-YSO03 C is the least evolved and thus the least massive.

\subsection{N113-YSO04}

N113-YSO04 exhibits two continuum sources, one compact and one extended and more diffuse. 
It is likely that most if not all of the H$_2$ emission detected in the FOV is ambient emission.

As in N113-YSO01, all of the emission in N113-YSO04 A appears to be either ambient emission (H$_2$ emission) or contaminating emission 
(Br$\gamma$ and He {\sc i}) from N113-YSO04 B.
Therefore the possibility that N113-YSO04 A is not a YSO cannot be excluded based on currently available data.
N113-YSO04 A exhibits a considerably redder continuum than N113-YSO04 B (see Fig. A1) and a significantly higher A$_V$, indicating a more embedded object.

The extended source (N113-YSO04 B) shows strong Br$\gamma$ and He {\sc i} emission which is slightly offset from the continuum
source. The underlying cause for the offset in the extended line emission is unclear but it is likely to be a result of the 
geometry of the central source and its projection onto the sky.
It also exhibits a similar expanding velocity profile for the Br$\gamma$ and He {\sc i} emission as N113-YSO03 A, implying that strong winds are present 
around both of these sources and that N113-YSO04 B is in a similar evolutionary state to N113-YSO03 A.
The detection of the He\,{\sc i} 2.113 $\mu$m doublet towards both sources is also consistent with an energetic, high density region such as a compact H\,{\sc ii}
region.

An H$_2$O maser has been previously identified to the North East of this region (See Fig. 1; \citealt{Imai2013}) however it has a relatively large separation of
9$\rlap{.}^{\prime\prime}$6$\pm$0$\rlap{.}^{\prime\prime}$5 (2.3$\pm$0.1 pc) which would indicate that this maser is not likely to be associated with 
either of the N113-YSO04 sources.

\section{Summary and Conclusions}

Using the SINFONI integral field spectrograph at the VLT, we have observed a sample of three 
\textit{Spitzer} selected YSOs in the bright dusty lane in N113.
The three targets look very similar at longer wavelengths:
all are classified as P- or PE-type sources ({\it Spitzer}/IRS spectrum rich in PAH emission and fine structure emission; \citealt{Seale2009}).
When the sources are observed at higher resolution in the \textit{K}-band, a wide variety of morphological
and spectral features is revealed. Our results are summarised below:
\begin{itemize}
\item Of the three \textit{Spitzer} sources, six distinct \textit{K}-band continuum sources have been resolved. N113-YSO01
contains only a single continuum source, N113-YSO03 contains three and N113-YSO04 contains two.
 \item Two sources (N113-YSO03 A, N113-YSO04 B) exhibit strong, extended wind features and are therefore likely to be compact H\,{\sc ii} regions, i.e. massive objects in
the final stages of star formation. The presence of the 2.113 $\mu$m He\,{\sc i} emission suggests that these
are indeed compact H\,{\sc ii} regions since in the C13 Galactic sample this doublet is only found towards H\,{\sc ii} regions.
\item N113-YSO03 C is dominated by H$_2$ emission, which is likely to occur in the collimated outflows driven by the
youngest hot core phase objects.
\item N113-YSO03 B appears to be a fairly typical, massive YSO: a point source with strong Br$\gamma$ emission and weaker He {\sc i} emission.
\item The remaining sources, N113-YSO01 and N113-YSO04 A, are compact and do not appear to have any emission lines associated with the continuum source
(all observed line emission appears to be ambient).
For N113-YSO04 A this emission is likely to be sourced from the extended source N113-YSO04 B whereas in N113-YSO01 the source of the line emission
appears to be outside the FOV.
Without additional data it is unclear where either of these sources would fall in an evolutionary context:
these objects could be featureless YSOs but the possibility of a non-YSO classification cannot be excluded.
\item Levels of extinction have been found to be typically lower than those within our Galaxy. The 
average extinction, $A_V$ and standard deviation of our sample is 22.3$\pm$9.3 mag compared with the values of
C13 of 45.7$\pm$17.6 mag. The average extinction towards massive YSOs in N113 is approximately half that of the Milky Way, 
consistent with the lower dust-to-gas ratio observed in the LMC \citep{Bernard2008} and an LMC metallicity of approximately half solar metallicity.
\item A number of interstellar H$_2$O masers and a single OH maser have previously been detected in the region;
however many of these in fact fall at a significant distance from the continuum sources ($>0.5$ pc) and are therefore unlikely to be excited by the 
sources resolved here. The one exception is N113-YSO03 C which is likely to be the excitation source of the water maser to the South East and also possibly 
the OH maser. 
\item Emission line fluxes are similar to those found in the Milky Way but the detection rates
of the He {\sc i} 2.058 $\mu$m emission line are higher in this sample than in C13. 
However
this is only a small sample and it may include a higher proportion of later stage YSOs and compact 
H\,{\sc ii} regions compared to C13.
\item The equivalent accretion luminosities calculated are consistent with the Galactic distribution but
appear to be high for their \textit{K}-band magnitudes.
The approach used here however does neglect any effects of metallicity \textit{K}-band continuum emission.
\item The CO bandhead, often associated with accretion discs in YSOs, is not 
detected in any of our sources. Even though the disc geometry can have a significant impact on the detection of CO \citep{Kraus2000, Barbosa2003}
 and the sample is small, 
the lack of CO bandhead emission could be due to CO destruction in the harder radiation fields associated with lower metallicity environments.
\item Finally we have mapped the velocity fields of extended gas structures around two of the 
continuum sources (N113-YSO03 A and N113-YSO04 B). These measurements imply stellar feedback in the form of stellar winds from 
newly formed massive stars at the centre of compact H\,{\sc ii} regions, driving expansion. 
\end{itemize}

Through high resolution characterisation of three similar \textit{Spitzer} sources, we have revealed a wide range of 
morphological and spectral properties.
Two out of the three \textit{Spitzer} sources in this 
study contain multiple YSOs and it is likely that this is not an uncommon occurance in the 
Magellanic Clouds. 
Crucially, we have shown that whilst \textit{Spitzer} and \textit{Herschel} observations provide valuable insights into the 
star formation process in the Magellanic Clouds, higher spatial resolution is required in order 
to develop a full understanding of the YSOs in question, especially when comparing them to Galactic samples.

\section*{Acknowledgments}
The authors thank the anonymous referee for his/her useful comments.
JLW acknowledges financial support from the Science and Technology Facilities Council of the UK (STFC) via the PhD studentship programme.
We would like to thank the staff at ESO's Paranal observatory for their support during the observations.
This paper made use of information from the Red MSX Source survey database at http://rms.leeds.ac.uk/cgi-bin/public/RMS\_DATABASE.cgi 
which was constructed with support from STFC. 
This research has made use of the SIMBAD data base, operated at CDS, Strasbourg, France.

\bibliographystyle{mn2e}
\bibliography{bibliography}

\appendix

\section{Extracted Spectra}

Figure A1 shows the 1D spectra extracted from the regions shown in Fig. 2 for all of the sources discussed in this paper.
The emission line identifications are marked on the spectra of N113-YSO03 C and N113-YSO01 with dotted lines showing the positions of the measured lines
in all spectra. The positions of all significant sky emission line residuals are marked on the spectrum of N113-YSO03 C in Fig. A2.
Figure A3 shows the positions of the Pfund series emission in the spectrum of N113-YSO03 A.

\begin{figure*}
\begin{minipage}{160mm}
\begin{center}
 \includegraphics[width=0.96\linewidth]{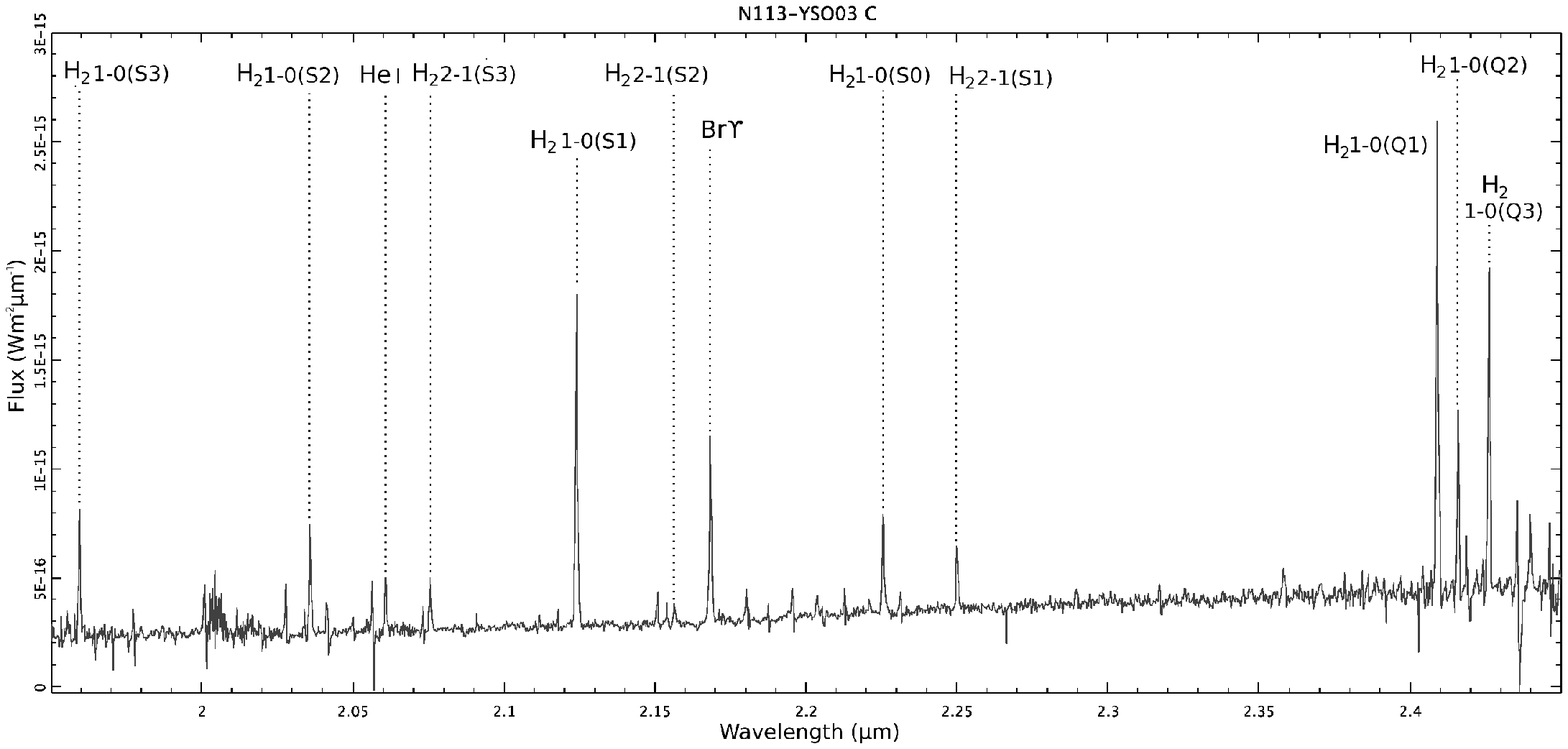}
 \includegraphics[width=0.96\linewidth]{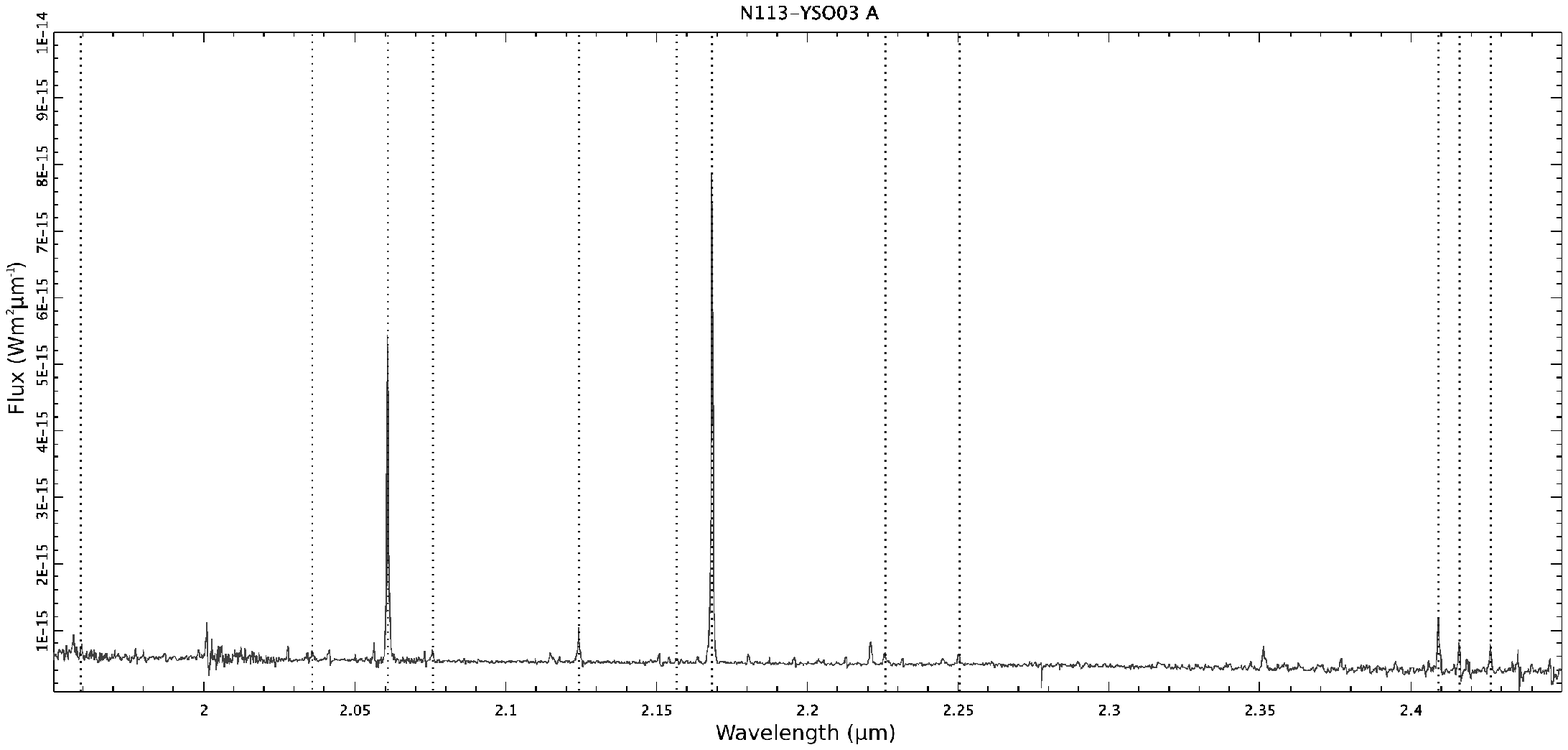}
 \includegraphics[width=0.96\linewidth]{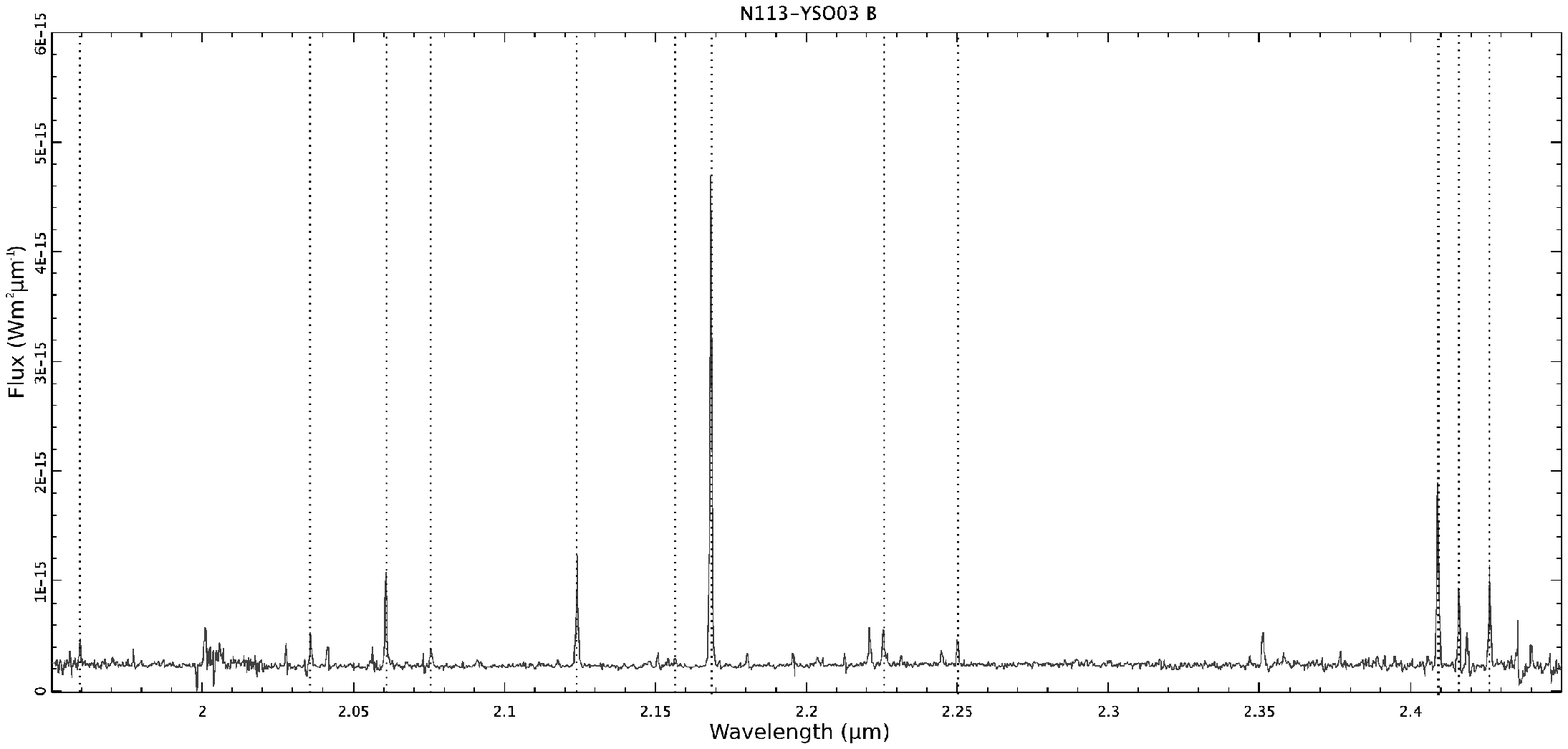}
\caption{Spectra of all N113 continuum sources extracted from the regions shown in Fig. 2. The spectrum of N113-YSO03 C (shown first) is marked with the 
positions of all of the measured spectral lines.}
\end{center}
\end{minipage}
\end{figure*}
\begin{figure*}
\begin{minipage}{160mm}
\begin{center}
 \includegraphics[width=0.96\linewidth]{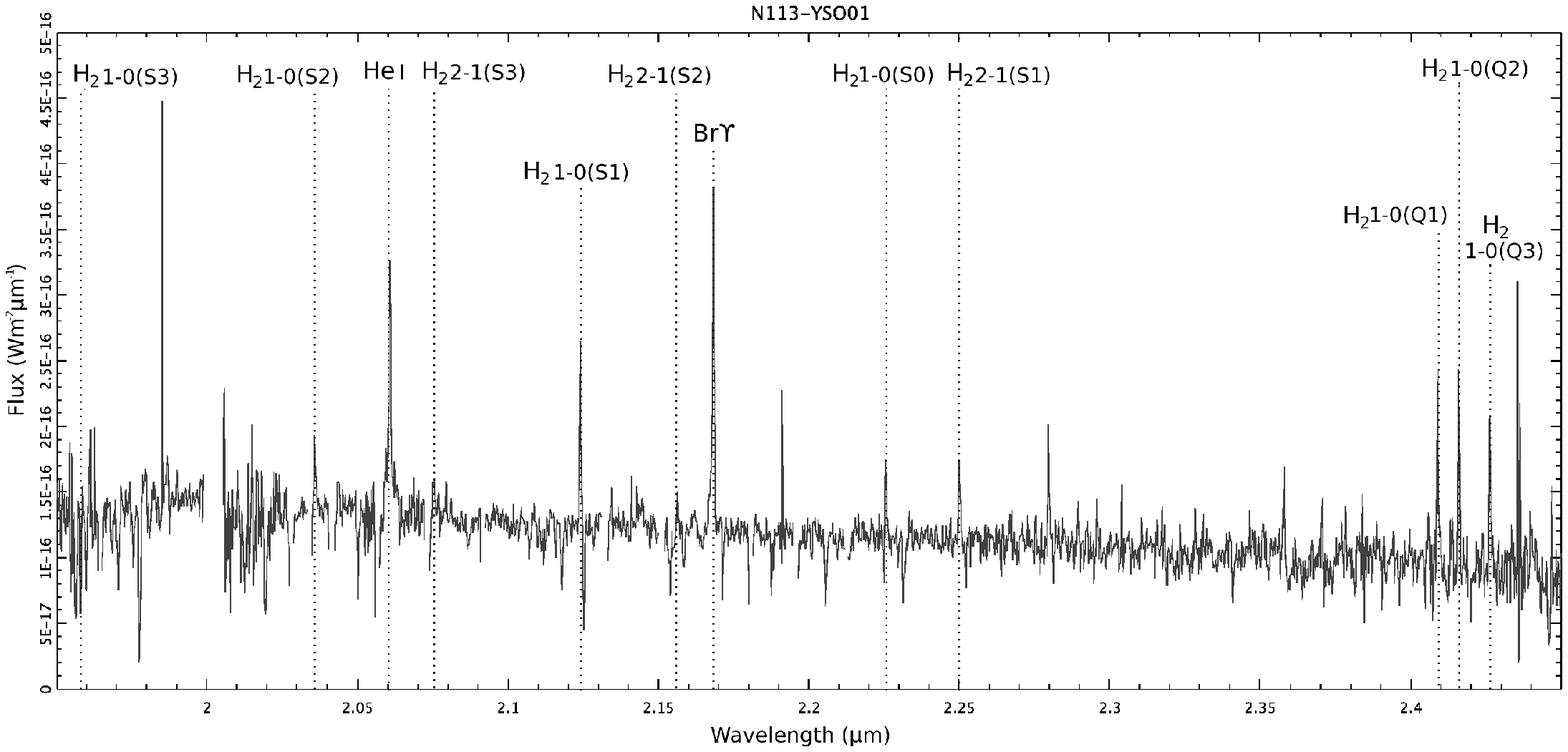}
 \includegraphics[width=0.96\linewidth]{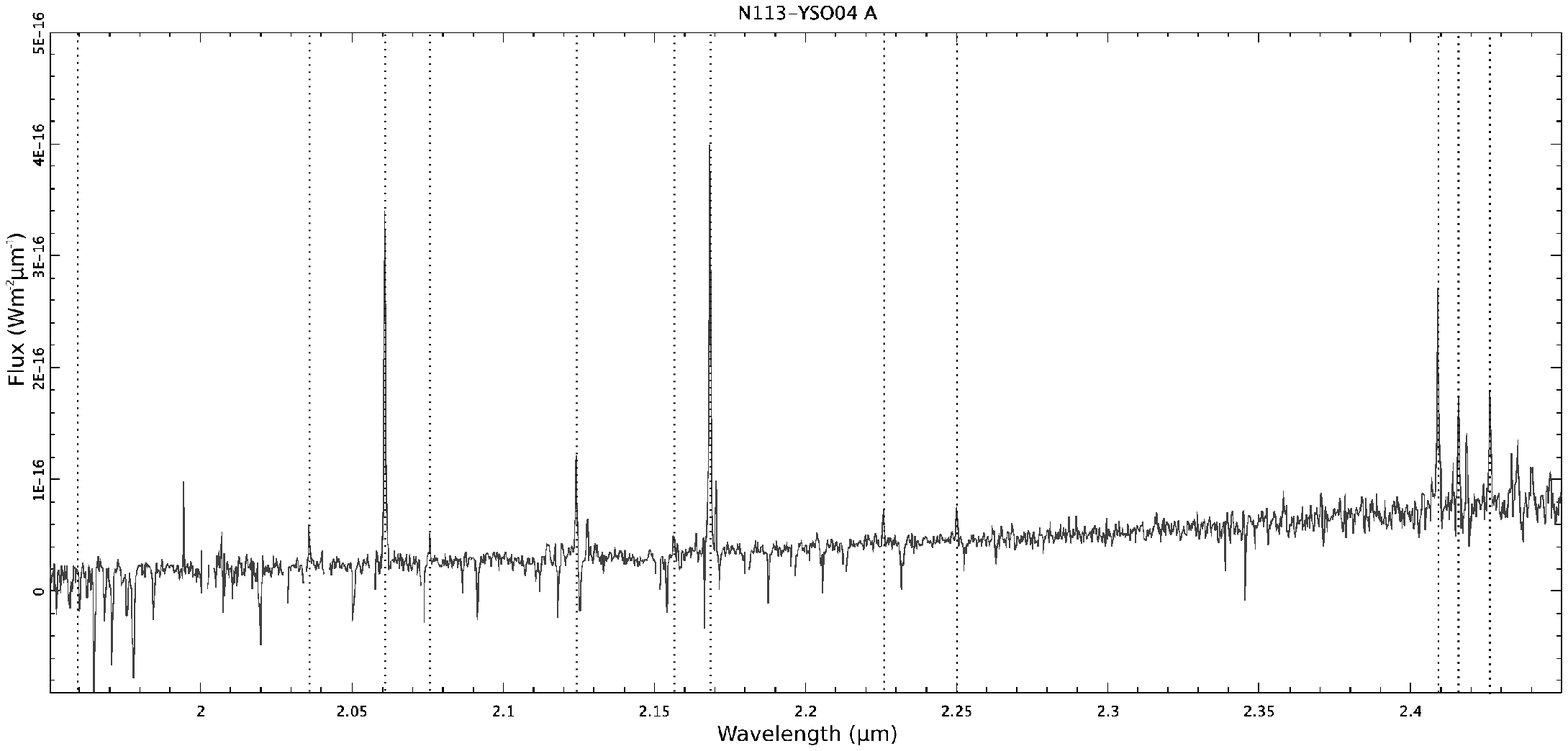}
 \includegraphics[width=0.96\linewidth]{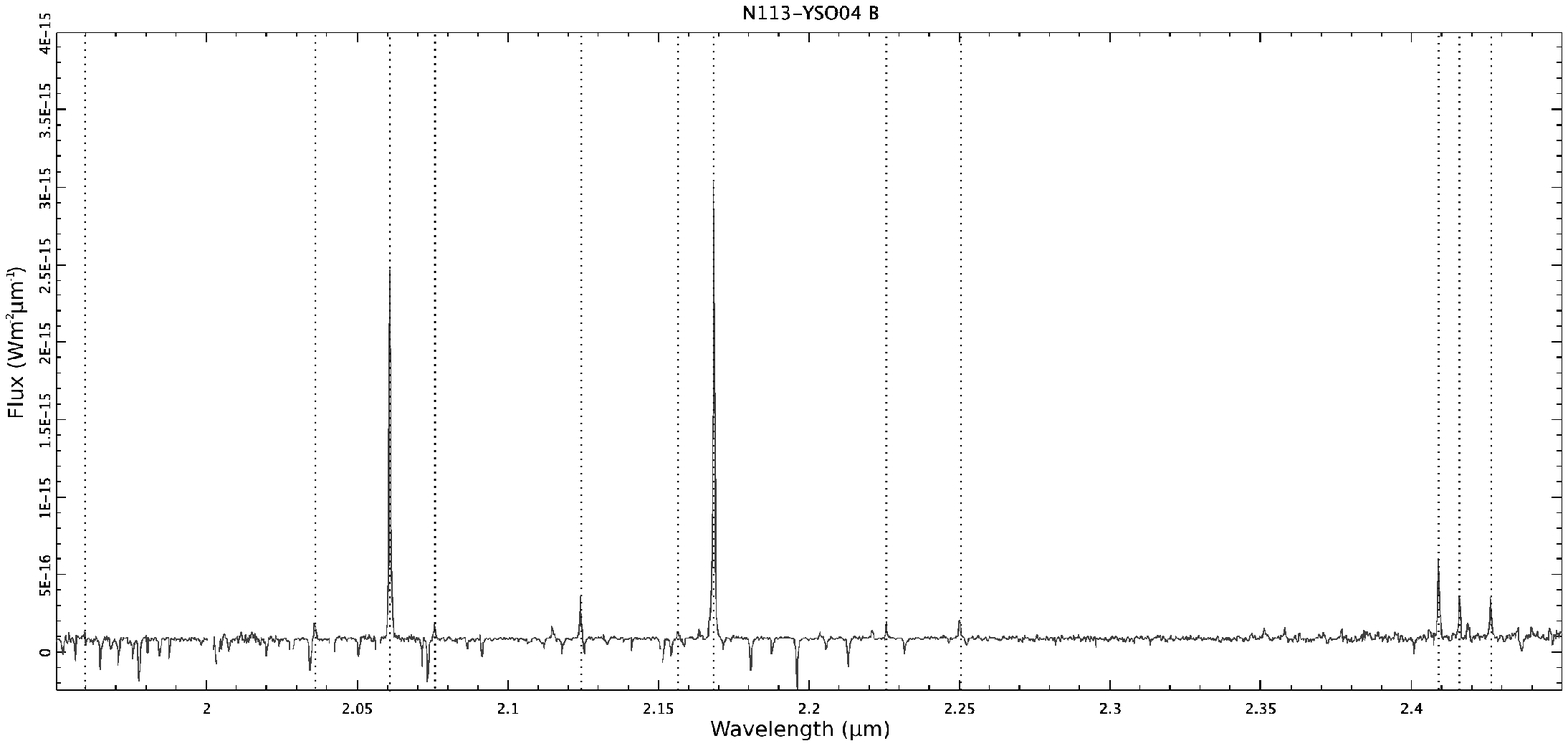}
\end{center}
\textbf{Figure A1 cont.} The spectrum of N113-YSO01 shows line identifications.
\end{minipage}
\end{figure*}

\begin{figure*}
\begin{minipage}{160mm}
 \includegraphics[width=0.98\linewidth]{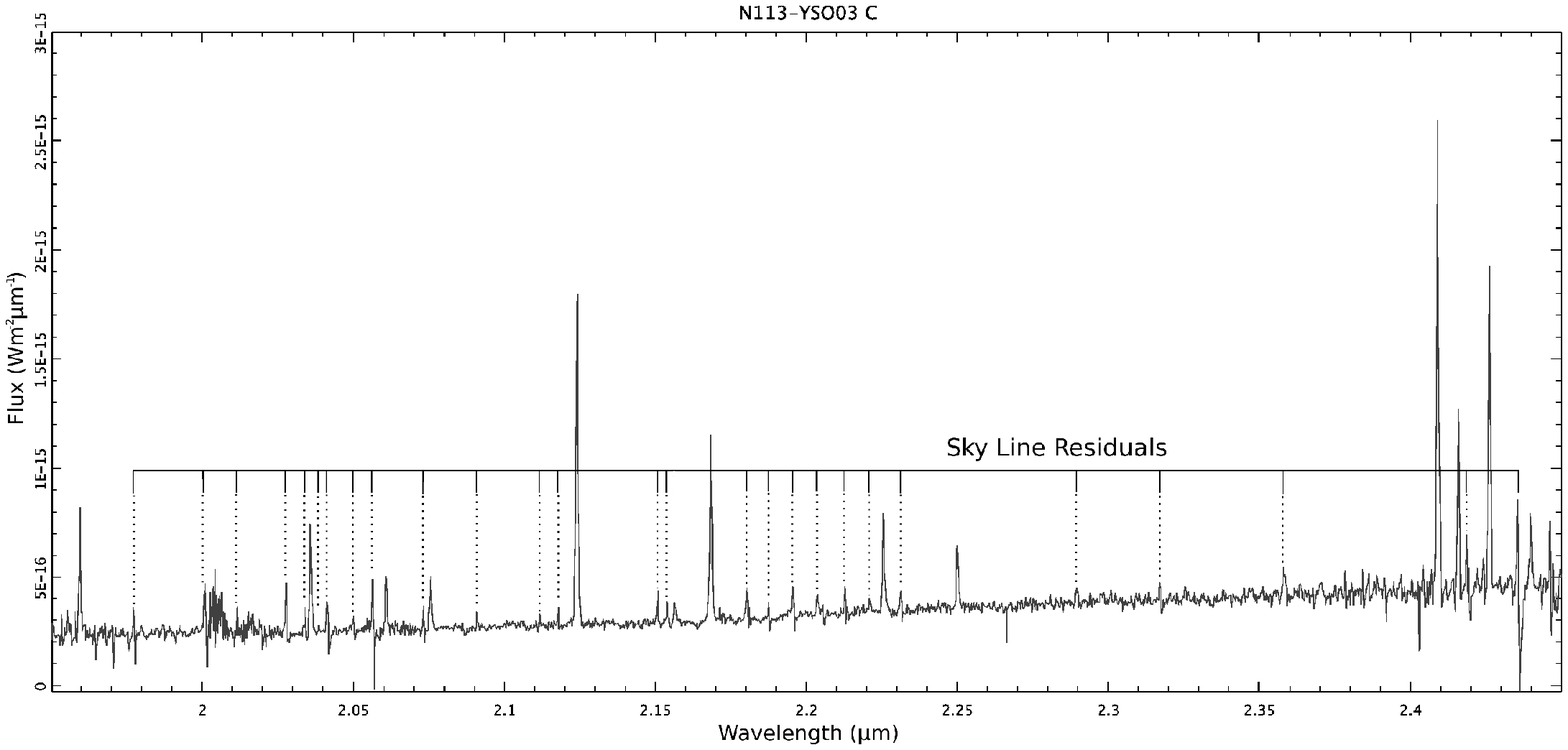}
\caption{Spectrum of N113-YSO03 C showing the positions of all of the sky line residuals.}
\end{minipage}
\end{figure*}

\begin{figure*}
\begin{minipage}{160mm}
 \includegraphics[width=0.98\linewidth]{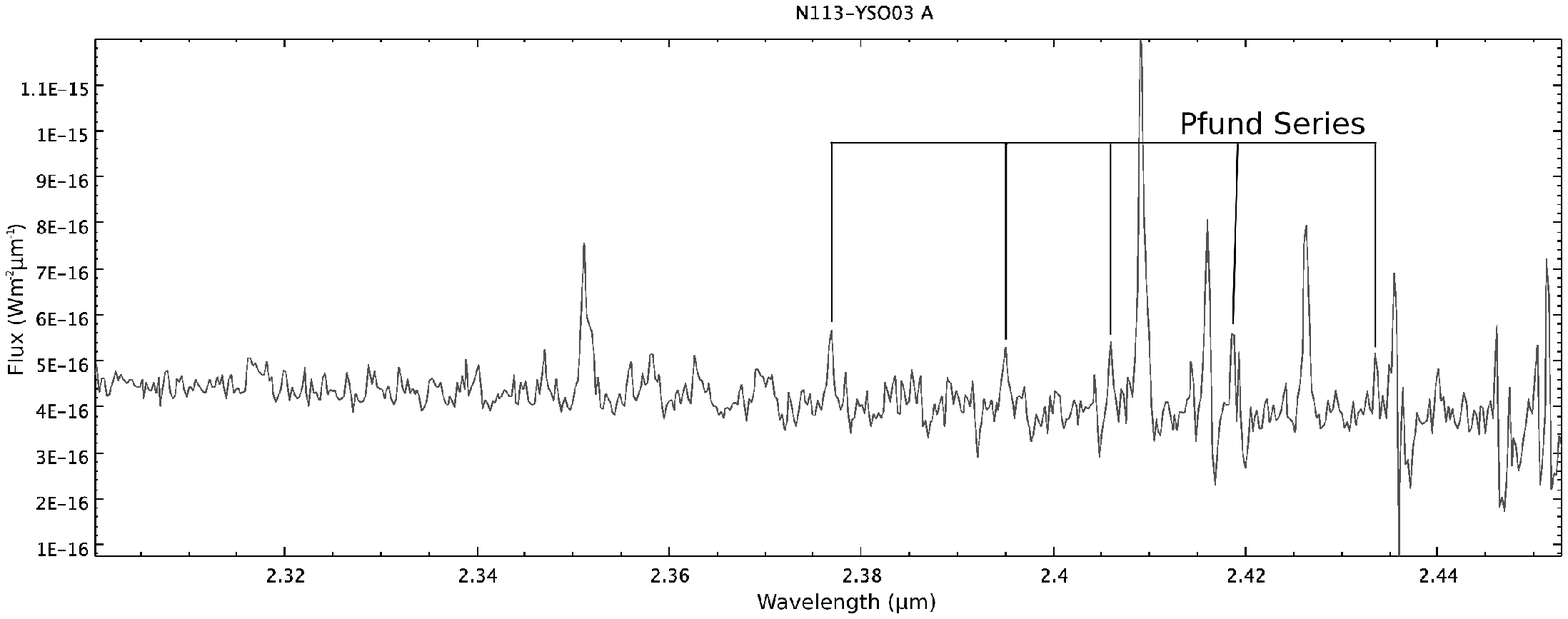}
\caption{Spectrum of N113-YSO03 A showing the positions of the Pfund series emission lines.}
\end{minipage}
\end{figure*}
\onecolumn
\section{Emission line fluxes}

Here we present the complete set of measured emission line fluxes obtained and analysed in Section 4.3. These fluxes have not been
correct for extinction and have been obtained by fitting Gaussian profiles to 1D spectra extracted from the regions shown in Fig. 2.

\begin{table}
 
\textbf{Table B1:} Emission line fluxes towards all newly resolved sources. No extinction correction has been applied. Where p appears in place of 
a flux this denotes that the line is present but the flux cannot be measured. \\
 \begin{tabular}{l l l l l l l}
\hline
  Object & He {\sc i} (2.058 $\mu$m) & Br$\gamma$ &  H$_2$ 1-0(S0) & H$_2$ 1-0(S1) & H$_2$ 1-0(S2) & H$_2$1-0(S3) \\
 & $10^{-19}$Wm$^{-2}$ & $10^{-19}$Wm$^{-2}$ & $10^{-19}$Wm$^{-2}$ &  $10^{-19}$Wm$^{-2}$ & $10^{-19}$Wm$^{-2}$ & $10^{-19}$Wm$^{-2}$   \\
\hline
N113-YSO01   & 1.92$\pm$0.25  & 2.00$\pm$0.13&  0.27$\pm$0.26  & 0.73$\pm$0.18  & 0.41$\pm$0.30 & \null \\
N113-YSO03 A & 37.4$\pm$2.2  & 53.8$\pm$2.4  & 1.67$\pm$0.30 & 3.66$\pm$0.50  & 1.19$\pm$0.20 & 1.30$\pm$0.41\\
N113-YSO03 B & 6.56$\pm$0.35  & 33.8$\pm$1.4 & 3.02$\pm$0.49 &  7.18$\pm$0.49  & 1.92$\pm$0.34 & 1.37$\pm$0.34\\
N113-YSO03 C & 1.53$\pm$0.20  & 7.70$\pm$0.38 & 3.64$\pm$0.30 & 11.7$\pm$6.1  & 3.47$\pm$0.36 & 4.09$\pm$0.50\\
N113-YSO04 A & 2.22$\pm$0.15 & 2.66$\pm$0.19 &  0.20$\pm$0.03 & 0.56$\pm$0.08  & 0.19$\pm$0.05 & \null \\
N113-YSO04 B & 17.1$\pm$1.0 & 21.1$\pm$1.4 &  0.56$\pm$0.12 & 1.74$\pm$0.17  & 0.75$\pm$0.24 & \null \\
\hline
Object & H$_2$ 2-1(S1) & H$_2$ 2-1(S2) & H$_2$ 2-1(S3) & 1-0Q(1) & 1-0Q(2) & 1-0Q(3) \\
 & $10^{-19}$Wm$^{-2}$ & $10^{-20}$Wm$^{-2}$ & $10^{-19}$Wm$^{-2}$ &  $10^{-19}$Wm$^{-2}$ & $10^{-19}$Wm$^{-2}$ & $10^{-19}$Wm$^{-2}$  \\
\hline
N113-YSO01 & 0.36$\pm$0.08 & 1.36$\pm$0.63 & 	\null  &0.94$\pm$0.18  & 1.09$\pm$0.10  & 0.83$\pm$0.13 \\
N113-YSO03 A  & 1.25$\pm$0.18 & 5.53$\pm$1.29 & 1.18$\pm$0.22 & 7.47$\pm$0.87  & 3.14$\pm$0.57  & 3.39$\pm$0.27  \\
N113-YSO03 B  & 1.38$\pm$0.14 & 4.65$\pm$0.97  & 1.42$\pm$0.16 & 12.2$\pm$1.0  & 5.34$\pm$0.44  & 6.93$\pm$0.54  \\
N113-YSO03 C  & 2.04$\pm$0.21 & 6.67$\pm$0.85  & 1.85$\pm$0.14 & 16.5$\pm$1.4  & 6.33$\pm$0.56  & 11.3$\pm$0.6  \\
N113-YSO04 A & 0.16$\pm$0.02 & 0.93$\pm$0.27  & 0.11$\pm$0.03 & 1.47$\pm$0.19  & 0.69$\pm$0.15  & 0.85$\pm$0.07  \\
N113-YSO04 B & 0.78$\pm$0.11 & 3.46$\pm$1.00  & 0.62$\pm$0.09  & 3.72$\pm$0.45  & 2.07$\pm$0.19  & 2.11$\pm$0.16  \\
\hline
Object & He\,{\sc i} (2.113 $\mu$m) & Pf 20-5 & Pf 21-5 & Pf 22-5 & Pf 23-5 & Pf 25-5 \\
 & $10^{-20}$Wm$^{-2}$ & $10^{-20}$Wm$^{-2}$ & $10^{-20}$Wm$^{-2}$ &  $10^{-20}$Wm$^{-2}$ & $10^{-20}$Wm$^{-2}$ & $10^{-20}$Wm$^{-2}$  \\
\hline
N113-YSO01 & \null & \null & \null & \null & \null & \null \\
N113-YSO03 A & 9.1$\pm$1.6 & 22.1$\pm$8.5 & p & 9.4$\pm$2.0 & 9.3$\pm$1.7 & \llap{1}0.1$\pm$1.3 \\
N113-YSO03 B & \null & \null & \llap{1}3.9$\pm$4.3 & & 4.8$\pm$1.2 & 6.7$\pm$1.0 \\
N113-YSO03 C & \null & \null & \null & \null & \null & \null \\
N113-YSO04 A & 2.4$\pm$0.5 & \null & \null & \null & \null & \null \\
N113-YSO04 B & 5.8$\pm$1.1 & \null & 8.1$\pm$2.4 & 5.3$\pm$1.3 & 5.2$\pm$1.9 & 3.6$\pm$0.9 \\
\hline
 \end{tabular}
\end{table}

\onecolumn
\section{Sky line velocity maps}

Here we present emission line centroid velocity maps for sky line emission as measured from the sky offset for N113-YSO03. 
The emission lines measured were chosen because they are relatively strong and cover a relatively large wavelength baseline encompassing all of the 
emission lines for which we have measured velocity profiles in Section 4.3. Within uncertainties there are no systematic velocity shifts across the detector.

\vspace{5mm}

 \begin{minipage}{170mm}
  \begin{center}
\includegraphics[width=0.48\linewidth]{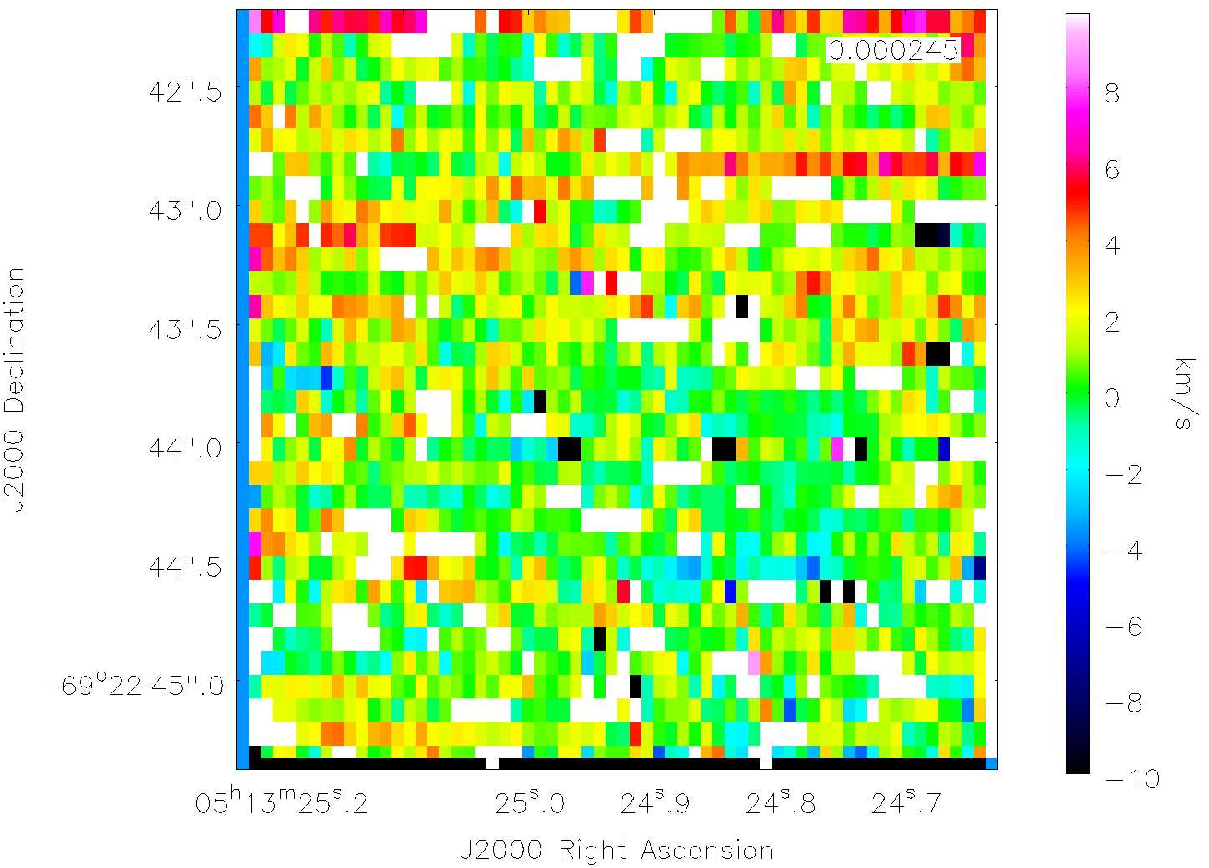}
\includegraphics[width=0.48\linewidth]{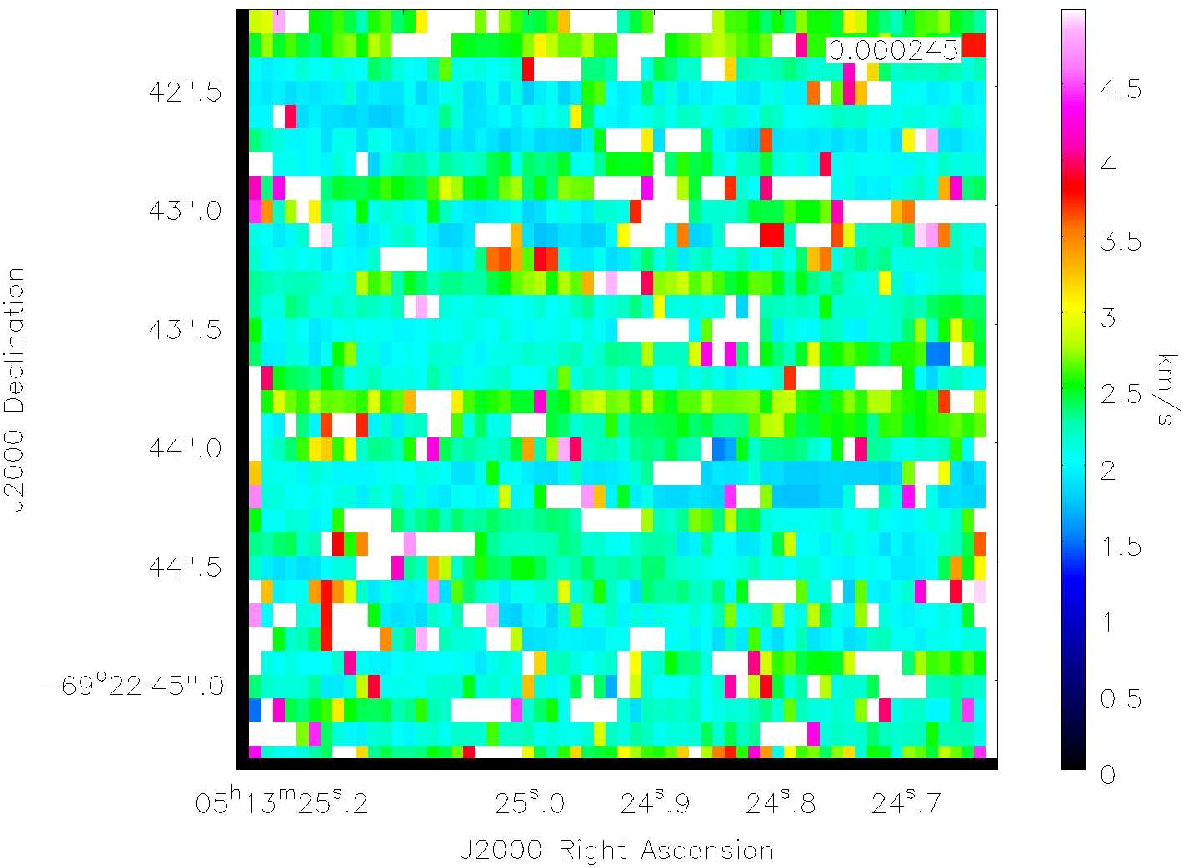}
\includegraphics[width=0.48\linewidth]{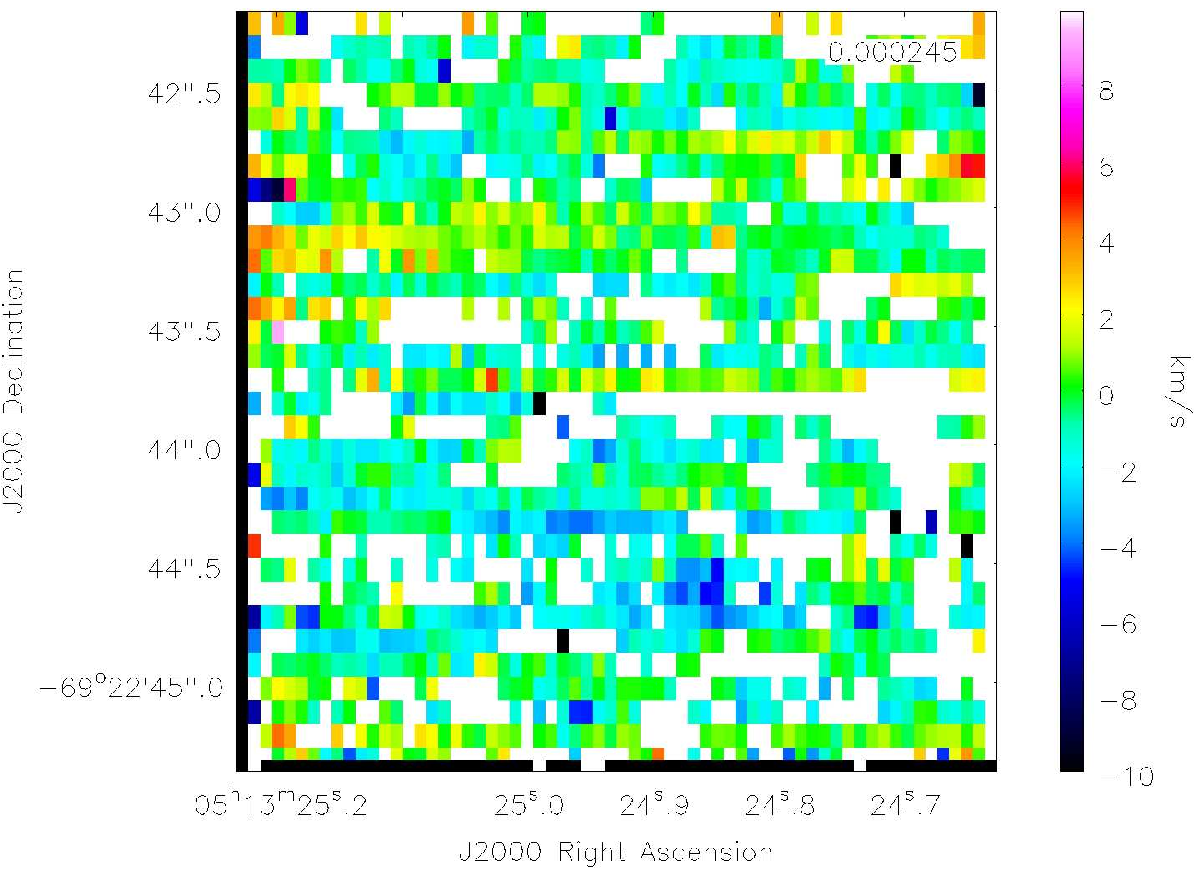}
\includegraphics[width=0.48\linewidth]{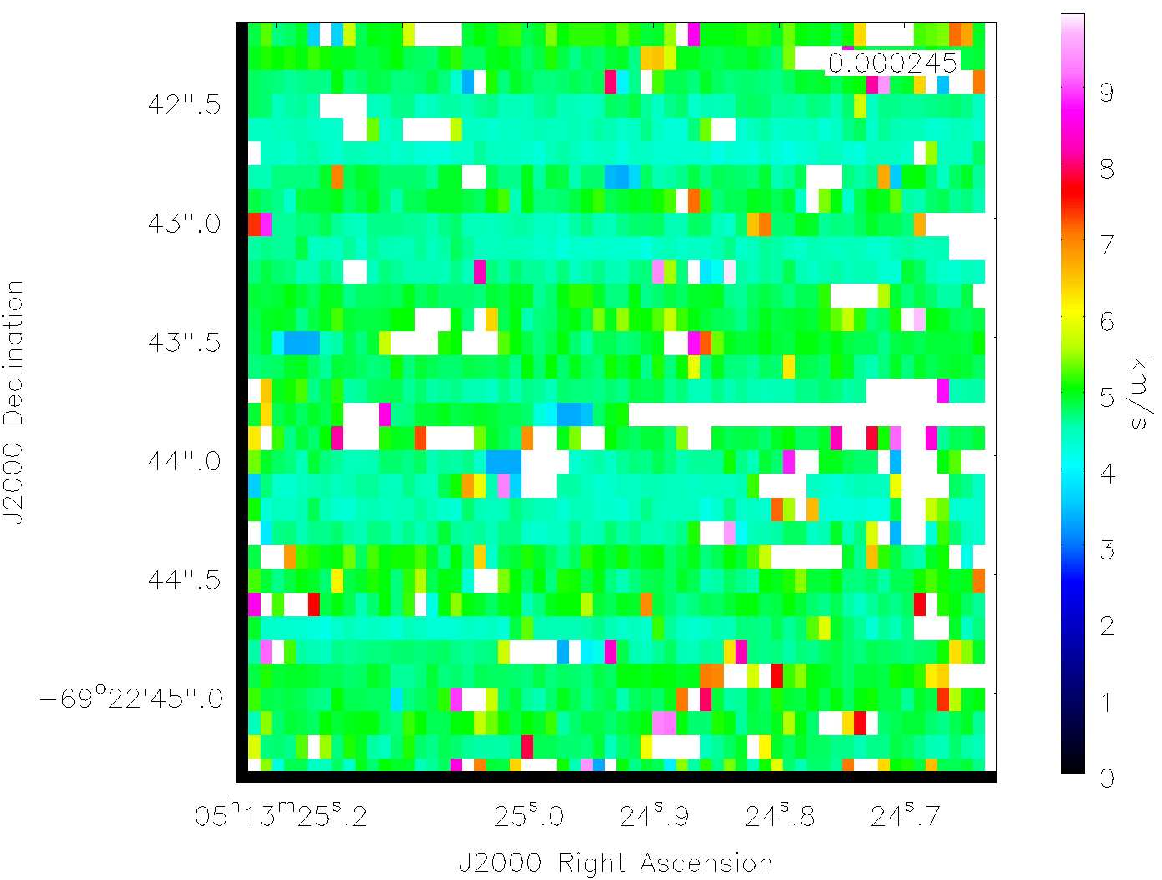}
\includegraphics[width=0.48\linewidth]{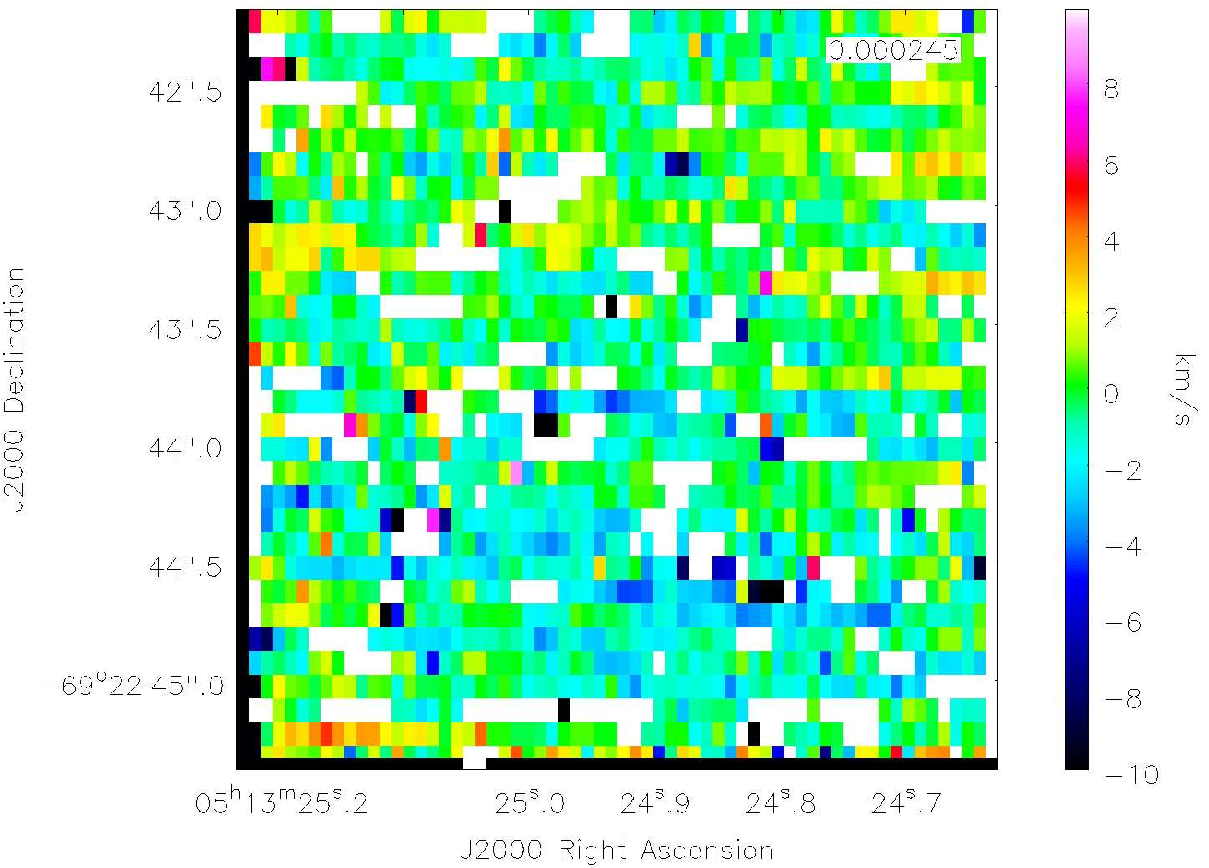}
\includegraphics[width=0.48\linewidth]{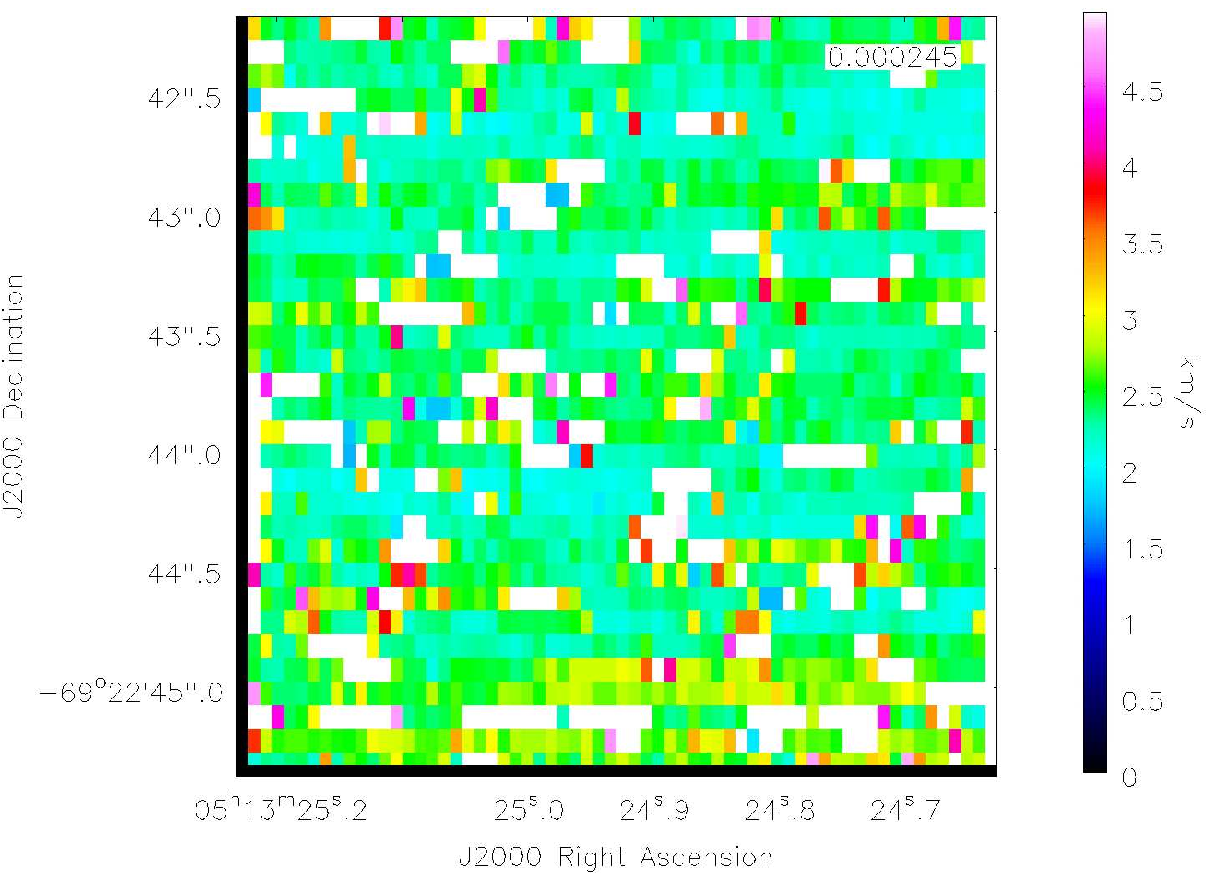}
 \end{center}
\textbf{Figure C1.} Relative velocity (left) and error maps (right) for sky emission lines in the N113-YSO03 sky cube. Top: 2.0567 $\mu$m; middle: 2.1511 $\mu$m;
bottom: 2.1806 $\mu$m. 
 \end{minipage}

 \label{lastpage}
\end{document}